\documentclass[useAMS,usenatbib]{mn2e}
\usepackage{times}

\usepackage{graphicx,color}

\input epsf

\usepackage{wrapfig}

\newcommand{\beq}{\begin{equation}}
\newcommand{\beqa}{\begin{eqnarray}}
\newcommand{\eeq}{\end{equation}}
\newcommand{\eeqa}{\end{eqnarray}}

\newcommand{\lsim}{\la}
\newcommand{\gsim}{\ga}

\newcommand{\vect}[1]{\mbox{\boldmath${#1}$}}
\newcommand{\lmk}{\left(}
\newcommand{\rmk}{\right)}
\newcommand{\lnk}{\left\{ }
\newcommand{\rnk}{\right\} }
\newcommand{\lkk}{\left[}
\newcommand{\rkk}{\right]}

\newcommand{\p}{\partial}

\newcommand{\ba}{\bar {\alpha}}
\newcommand{\bb}{\bar {\beta}}
\newcommand{\bd}{\bar {\delta}}
\newcommand{\be}{\bar {\epsilon}}
\newcommand{\dv}{\dot {\varpi}}

\title{Analyses on a Relativistic Hierarchical Resonance with the Hamiltonian Approach}

\author[N. Seto ]{Naoki Seto
\\
Department of Physics, Kyoto University, 
Kyoto 606-8502, Japan
}

\begin{document}

\maketitle

\begin{abstract}

We study dynamical evolution of a  resonant triple system formed by an inner EMRI 
and an additional outer MBH.  The relevant resonant state  
($\lambda_2-\varpi_1\sim const$)  is supported by the relativistic apsidal 
precession of the inner EMRI,
and, unlike standard mean motion resonances,  the triple system can have a 
hierarchical orbital configuration (but different from the Kozai process).  In 
order to 
analyze this unusual  resonant  system,  we extend the so-called Hamiltonian  
approach,  and derive a mapping  
from the EMRI-MBH triple system to a simple one-dimensional Hamiltonian.
 With the derived mapping,  we make analytical 
predictions for characteristic quantities of the resonance, such as the capture 
probability,   and find that they reasonably agree with 
numerical simulations up to moderate eccentricities.
\end{abstract}


\section{introduction}
In  the solar system, orbital resonances are broadly observed at various spatial 
scales (Peale 1986; Murray \& Dermott 2000 (hereafter MD)).  For example, Pluto and Neptune have orbital periods of 3:2 and their 
orbital stability is sustained by this simple relation. The resonant states with such 
commensurable orbital periods are termed mean motion resonances (MMRs), and have 
been identified also among extrasolar planetary systems \citep{2011ApJS..197....8L,2012arXiv1211.5603P}.

In a recent paper (Seto 2012),     triple system formed by  an EMRI  
 and an additional outer massive black hole (MBH)  
was 
numerically studied, using the
post-Newtonian (PN) approximation (see Fig.\ref{fig1} for the orbital configuration).  Here ``EMRI'' stands for ``extreme-mass-ratio 
inspiral'' and represents an inspiral of a compact object (CO) around a MBH (see \cite{2004CQGra..21S1595G} for detail).  The numerical simulations were performed  mainly from small initial 
eccentricities, and  two resonant 
states were identified with  $\lambda_2-\varpi_1\sim const$ and 
$3\lambda_2-\varpi_1-2\Omega_1\sim const$.  Here $\lambda_2$
 represents the mean anomaly of the outer MBH around the central MBH.  The 
 angles $\varpi_1$ and $\Omega_1$ are the longitudes of the pericenter and the 
 ascending node of the CO.

Seto (2012) also discussed astronomical aspects for the triples, including 
 prospects for gravitational wave  and electromagnetic wave observations.
The expected numbers of resonant captures (not the capture probabilities at the 
resonant encounters) were roughly estimated and the mode $\lambda_2-\varpi_1\sim 
const$ turned out to occupy the majority of the capture events.

The two resonant states are induced by the relativistic 
 apsidal precession of the EMRI and do not depend on the inner mean anomaly $\lambda_1$, 
 unlike the standard MMRs in which  two terms proportional to the inner and outer 
 mean anomalies nearly cancel  \citep{1986sate.conf..159P,2000ssd..book.....M}. 
 Consequently, the resonant EMRI-MBH system can have a 
 hierarchical orbital configuration and the masses of the two MBHs can be 
 comparable.  These properties are remarkably different from the standard MMRs 
 where two orbital periods (equivalently, two semimajor axes) are comparable but 
 the masses of the central body must be much larger than other ones due to 
 orbital stability \citep{1993Icar..106..247G}.

In this paper, we focus our analysis to the resonant dynamics of the dominant 
mode $\lambda_2-\varpi_1\sim const$, paying special attention to  dependence 
on the inner eccentricity. To this end, we utilize 
the so-called Hamiltonian approach in celestial mechanics \citep{1972MNRAS.160..169S,1979CeMec..19....3Y,1982CeMec..27....3H,1983CeMec..30..197H,1986sate.conf..159P,2000ssd..book.....M}. This method has been 
applied for the standard MMRs.  Its basic prescription is to extract the essential 
dynamical degree of freedom from the complicated original triple system and map the triple 
system down to a simple one-dimensional system 
whose dynamics is  described by a rescaled Hamiltonian  (more precisely, in a 
two-dimensional phase space with  a canonical
variable and its conjugate momentum).  {Our resonance is an unusual mean motion 
resonance, but certainly classified as an eccentricity-type resonance. Therefore, the
 important dynamical parameters would be  the inner eccentricity and the resonance 
 angle  $\lambda_2-\varpi_1\sim const$.  Around the resonance, other parameters 
 approximately behave as cyclic variables or constants (see {\it e.g.} MD).
}

So far, various characteristic 
behaviours of the standard MMRs have been  successfully explained with the 
Hamiltonian approach, taking advantage of  
basic principles on analytical mechanics, such as conservation of  adiabatic  
invariants \citep{1984CeMec..32..127B,1986sate.conf..159P,2000ssd..book.....M}.
In this paper, we are primarily interested in whether we can 
suitably extend the Hamiltonian approach for our unusual resonant state  
$\lambda_2-\varpi_1\sim const$.  If it works well, we can easily make 
astrophysical arguments on the resonant EMRI-MBH systems without using costly numerical 
simulations, and, furthermore, we can better understand the efficient analytical 
approach itself in a perspective different from the traditional analyses for the 
standard MMRs.

In  this paper, by appropriately handling the effects of the relativistic apsidal precession, 
 we derive the mapping from the EMRI-MBH triple system to the simple Hamiltonians 
 whose forms  are identical to those used for analyzing the standard MMRs.  We 
 then make analytical predictions on the dynamical evolution of the hierarchical 
 triples around the resonant encounters.  We compare these 
 predictions with numerical simulations and confirm good agreements for certain 
 range of the  eccentricity $e$ of the inner EMRI.

This paper is organized as follows. In \S 2
we summarize basic notations, briefly describe our numerical scheme, and provide 
some of representative numerical results around the resonant encounters.  In \S 
3 we discuss the relativistic apsidal precession. Later, its dependence  on the 
inner eccentricity $e$ plays a critical role for the overall structure of the 
mapping.  In \S 4, we compare the strengths of the first-order term ($\propto 
e\cos(\lambda_2-\varpi_1)$) and the second-order one ($\propto e^2 
\cos2(\lambda_2-\varpi_1))$ for our resonant state. In \S 5, we derive 
the mapping mentioned above, by  extending  the previous studies  done for 
the standard MMRs.  In the next three sections, using the derived mapping,  we make analytical predictions 
on the resonant dynamics and extensively compare them with numerical 
simulations.  The capture rate is examined in \S 6. In  \S 7, we discuss  the 
gap of 
 the eccentricity observed at a failure of resonant capture. In \S 8, we study 
 resonant encounters for relatively inclined orbits. We summarize this paper in \S 9.

\section{Evolution of the system}

\begin{figure}
\includegraphics[width=85mm]{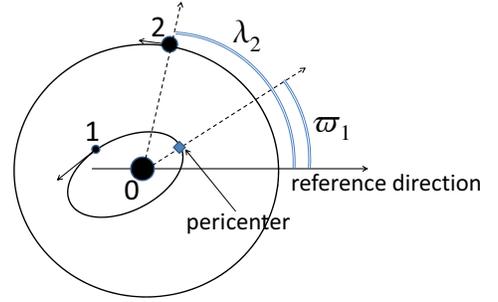}
\caption{ The coplanar triple system composed by two massive black holes (MBHs; 
0 and 2) and a compact object (CO; 1).  The MBH 0 and the CO 1 form an inner 
EMRI.   We measure the angular position $\varpi_1$ of the pericenter of the CO  with respect to the  fixed reference direction. The outer 
MBH has a nearly circular orbit and its angular position is given by 
$\lambda_2$.  We can similarly define the variable $\lambda_1$ for the CO, 
but it does not appear in our resonant variable $\lambda_2-\varpi_1\sim const$.} 
\label{fig1}
\end{figure}

Our triple system is composed by two MBHs with masses $m_0$, $m_2$ and a CO of $m_1 (\ll m_0,m_2)$.  The two components $m_0$ and 
$m_1$ form an inner EMRI and the third  one $m_2$ is  rotating outside 
the EMRI (see Fig.\ref{fig1}). For the 
orbital elements of the triple, we follow the positions of $m_1$ and $m_2$ 
relative to the central MBH $m_0$ and determine the (instantaneous) semimajor 
axes $a_l$ and eccentricities $e_l$ ($l=1,2$).  Since we only handle triples with 
nearly circular outer orbit and the outer eccentricity $e_2(\ll1)$  is not important in this paper, we put $e_1=e$ for simplicity 
 of notation.  Except for \S 8, 
we mainly study coplanar orbital configurations, as shown in Fig.\ref{fig1}, and define 
the mean anomalies $\lambda_l$ and the longitudes of pericenters $\varpi_l$ ($l=1,2$), 
following the  standard convention \citep{2000ssd..book.....M}.  Below, we use the
geometrical unit $G=c=M=1$ ($M\equiv m_0+m_1+m_2$: the total mass).

For numerical evolution of the system, we use the three-body ADM Hamiltonian $H_{TB}$ 
in the post-Newtonian formalism, and  neglect effects of spins.  The Hamiltonian is expanded as
\beq
H_{TB}=H_N+H_1+H_{2.5}\label{ham3}
\eeq
 \citep{1987PhLA..123..336S,1997PhRvD..55.4712J,2008CQGra..25s5019L,2009PhRvD..80l4018A,2010arXiv1012.4423G} (see also Moore 1993).
Here $H_N$ is the Newtonian term, and  $H_1$ is the 1PN term, namely  the leading order 
relativistic correction.  The 2.5PN term $H_{2.5}$ is the first dissipative 
term caused by gravitational radiation reaction, and invokes the orbital decay of 
the system. In Eq.(\ref{ham3}),  we put the subscript ``$TB$'' representing 
``three-body'' to distinguish the rescaled Hamiltonian $H$  introduced in \S 5.

 In the previous paper \citep{2012PhRvD..85f4037S}, we included the 2PN term $H_2$. But this term is time 
 consuming and less important for our resonance.  We 
 thus drop it here.

The equations of motions for the positions ${\vect x}_l$ and momenta ${\vect 
p}_l$ of the three masses $m_l$ ($l=0,1,2$) are obtained by taking appropriate 
partial derivatives of the Hamiltonian. As in Seto (2012), we use the new 
variable ${\vect s}_l\equiv {\vect p}_l/m_l$ to properly handle the motion of 
the CO with $m_1\ll 1$ (including the test particle limit $m_1=0$). These equations are 
integrated by a Runge-Kutta method with an adaptive step size control (Press et 
al. 1996, and see also Seto 
\& Muto
2011 for detail).

\begin{table}
\begin{tabular}{lcccccccl}
\hline\hline
\ case & $m_0$ &  $m_1$ &  $m_2$ & $a_{1in}$ & $\frac\alpha8$ &$ \frac{d\bd}{d\tau}$  & $D$ \\
\hline
\ I & 0.90 &  0& 0.1  & 50 &0.017 &0.00168 & 24.7\\
\ II & 0.98 & 0 & 0.02& 30 &0.027 &0.0289 & 68.0 \\
\ III & 0.999 & 0 & 0.001& 20& 0.035 &1.43  &890 \\
\hline\hline
\end{tabular}
\caption{The model parameters adopted in our numerical simulations. The initial outer semimajor axis $a_{2in}$ is adjusted to yield a 
resonant encounter. The  outer eccentricity is initially set at $e_{2in}\simeq 
0$ and  it stays 
at a small value. The left three columns show the basic parameters characterising 
  the resonant 
dynamics, and they are evaluated for coplanar orbits.  The ratio $\alpha/8$ is 
the critical eccentricity for 
the shift of the resonant angle $\lambda_2-\varpi_1$ (discussed in \S 4). The 
transverse velocity $d\bd/d\tau$ and the coefficient $D$ are defined in \S 5, and  evaluated with respect to the Hamiltonian (\ref{sh2}) for the second order 
resonance.
\label{tab:detectors}}
\end{table}

In Table 1, we present the model parameters of our numerical simulations. 
Since dependence of the resonant dynamics on the inner eccentricity $e$ is our
central issue,  we systematically analyze it for commonly arranged sets of 
parameters such as masses $(m_0,m_1,m_2)$ and the initial inner semimajor axis 
$a_{1in}$. Among the three models listed  in Table 1, 
we mainly use  
models I and II, targeting comparable MBHs, and   
model III is studied for a specific purpose in \S 4.

In Figs.2-4,   we present samples of typical orbital evolutions of model I around the 
resonant encounters.  We set 
 the initial outer distance $a_{2in}$ so that the system transverses the resonant condition due to the 
 radiational orbital decay.  Throughout this paper, we use the outer semimajor axis 
 $a_2$  to show the time.   This variable $a_2$ is  monotonically decreasing from its 
 initial value $a_{2in}$.

\begin{figure}
\includegraphics[width=40mm]{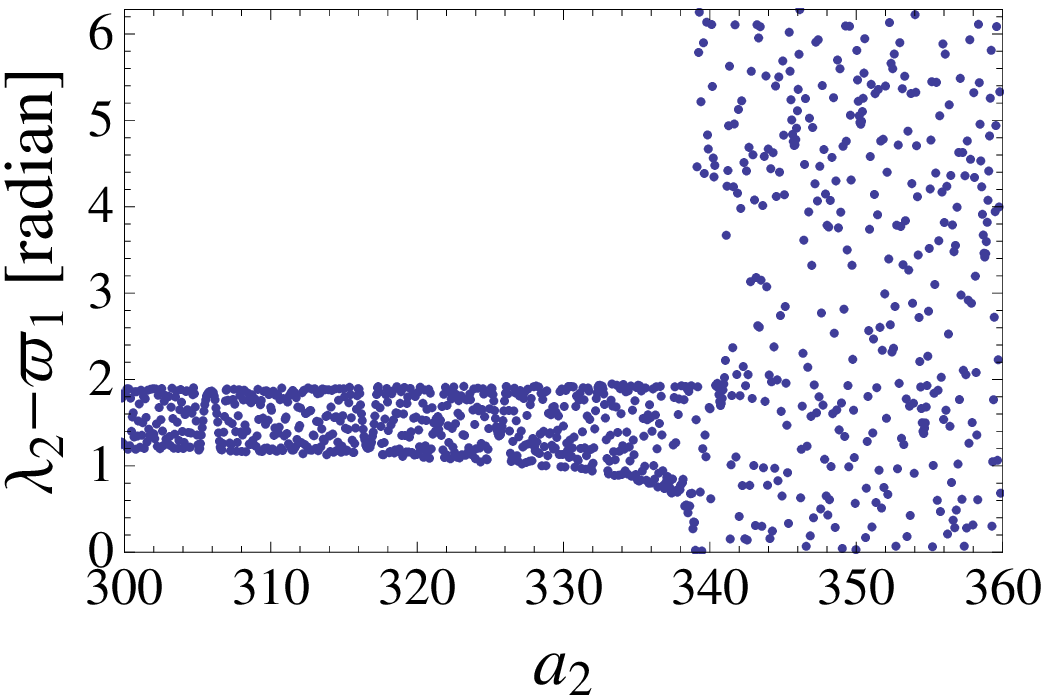}
\includegraphics[width=40mm]{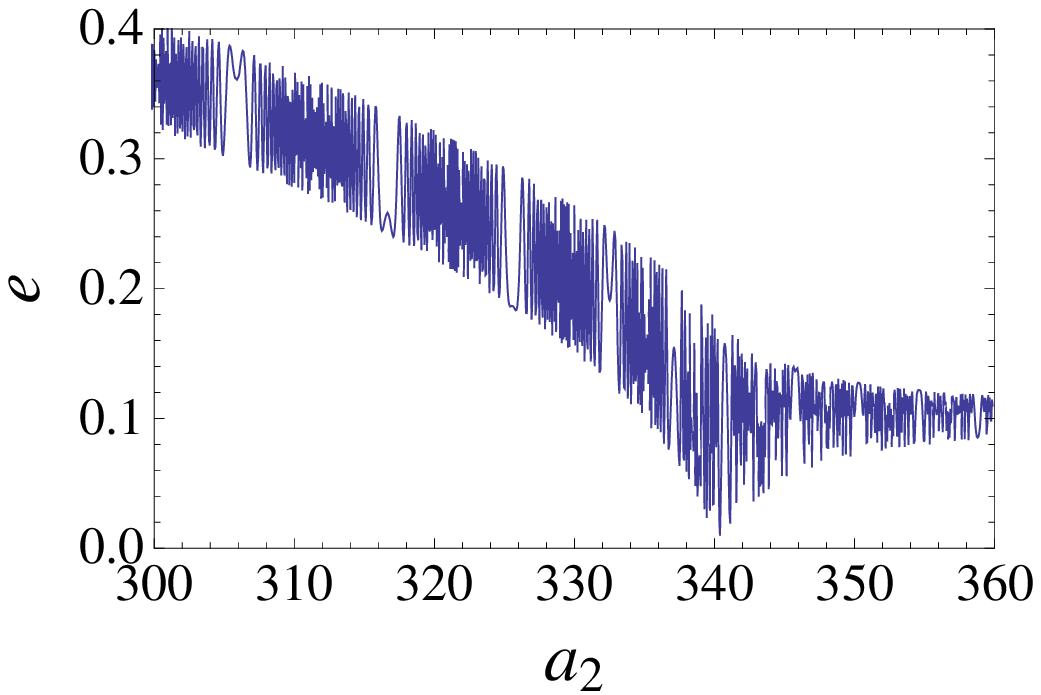}
\includegraphics[width=40mm]{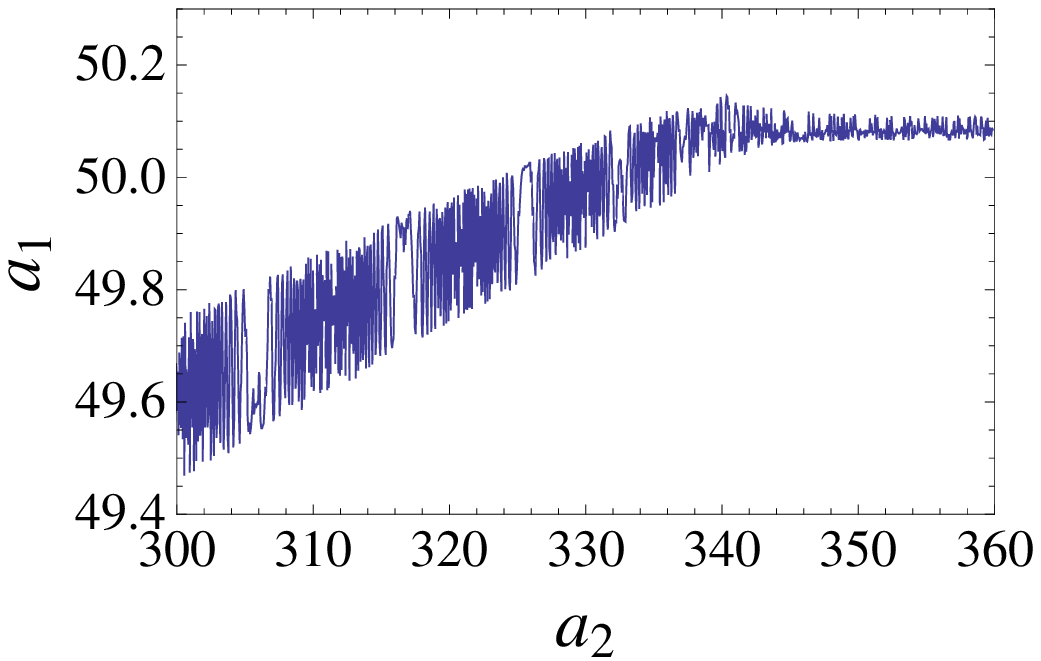}
\caption{Evolution of a coplanar EMRI-MBH triple system from an inner eccentricity 
$e\sim 
0.1$ (model I).   The horizontal axis $a_2$ is the semimajor 
axis of the outer MBH 
$m_2$ and decreases from its initial value $a_{2in}=360$ due to gravitational wave emission. The three panels show the 
resonant variable $\lambda_2-\varpi_1$  (the upper left one), the inner eccentricity 
$e$ (the upper right one, in modulo $2\pi$) and the inner semimajor axis $a_1$ (the bottom one). 
This system encounters the resonance around $a_2\simeq 340$, and the capture is 
successful. The inner eccentricity $e$ starts to increase afterward. 
}
\label{es}
\end{figure}

In Fig.\ref{es}, we show the results from an initial inner eccentricity $e_{in}\simeq 0.1$. 
The test particle $m_1$ is resonantly captured  by the outer MBH binary at the time $a_2\sim 340$,  
corresponding to the ratio of orbital periods at $\sim 17$. 
Incidentally, the inner eccentricity $e$ starts to grow and the inner axis $a_1$ 
decreases very slowly.   The resonant variable $\lambda_2-\varpi_1$ soon localizes around $\sim \pi/2$.

\begin{figure}
\includegraphics[width=40mm]{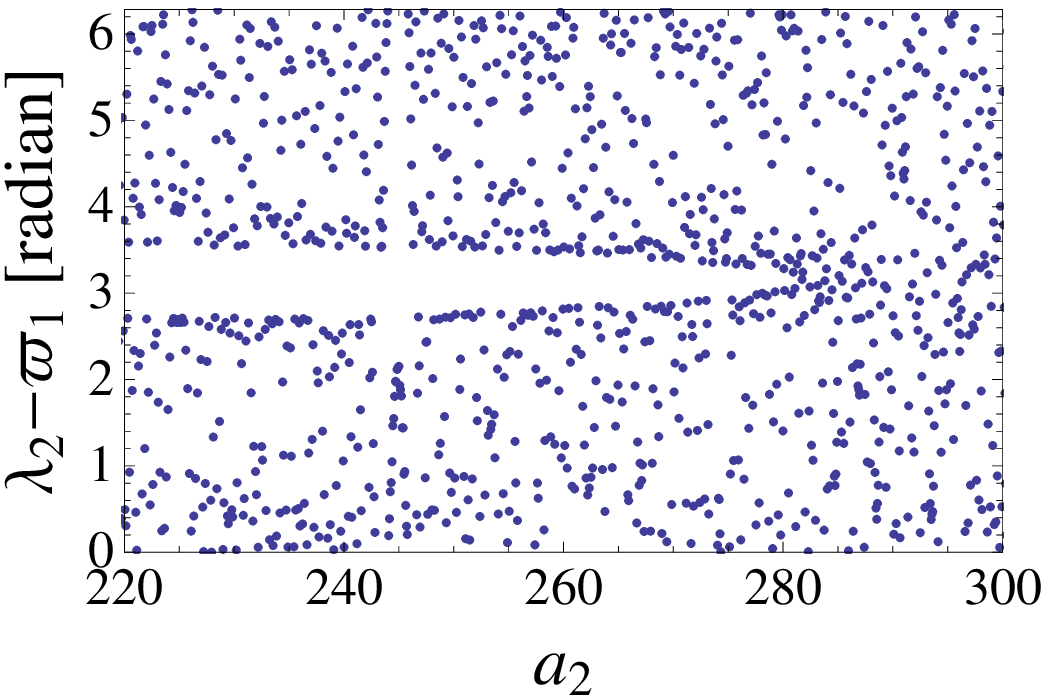}
\includegraphics[width=40mm]{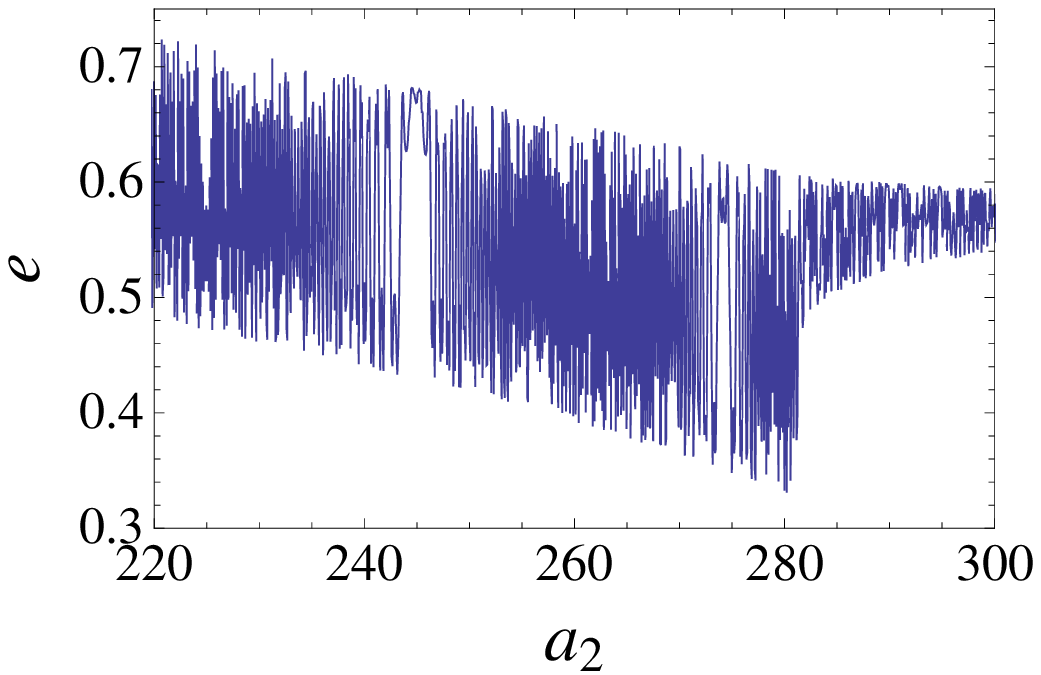}
\includegraphics[width=40mm]{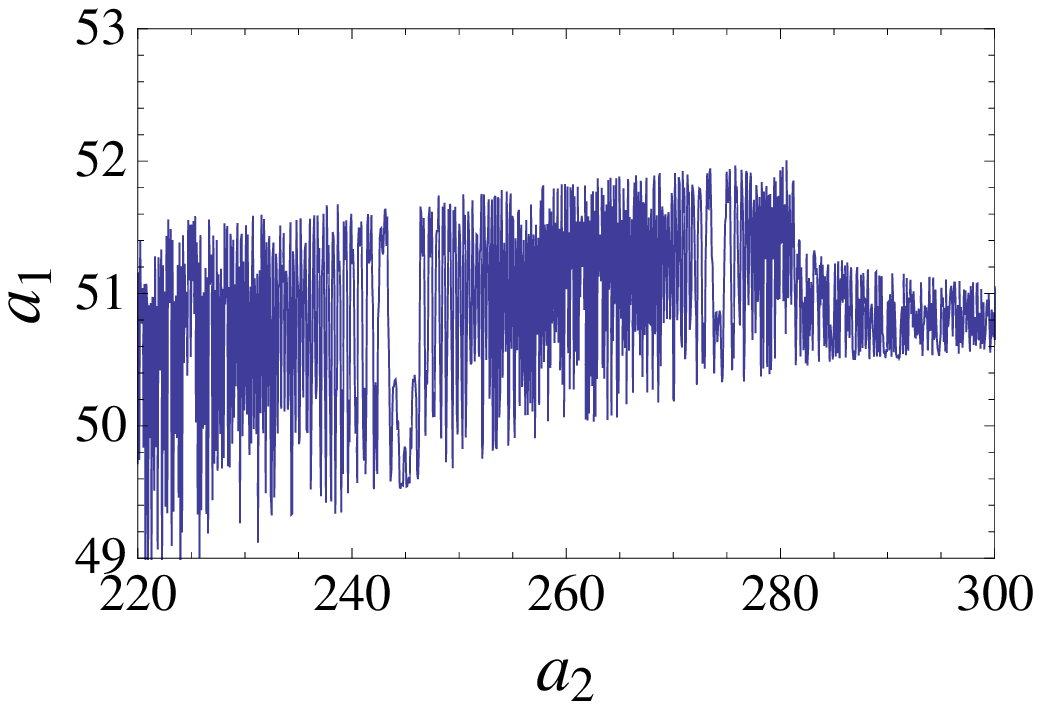}
\caption{Evolution of a coplanar triple from a large inner eccentricity $e\sim 0.57$   (model I). The inner EMRI is resonantly captured by 
the outer MBH around $a_2\sim 280$.  The libration amplitude is large.
}
\label{el1}
\end{figure}

For the run shown in Fig.\ref{el1}, we set  a larger initial eccentricity $e_{in}\simeq 
0.57$.  The test particle $m_1$ is captured into the resonance around $a_2\sim 280$.  In contrast 
to Fig.\ref{es}, the combination $\lambda_2-\varpi_1$ now has a large librational 
amplitude with a small excluded region around $\pi$.  Again, after the resonant capture, the inner eccentricity $e$ 
increases and the semimajor axis decreases.

\begin{figure}
\includegraphics[width=40mm]{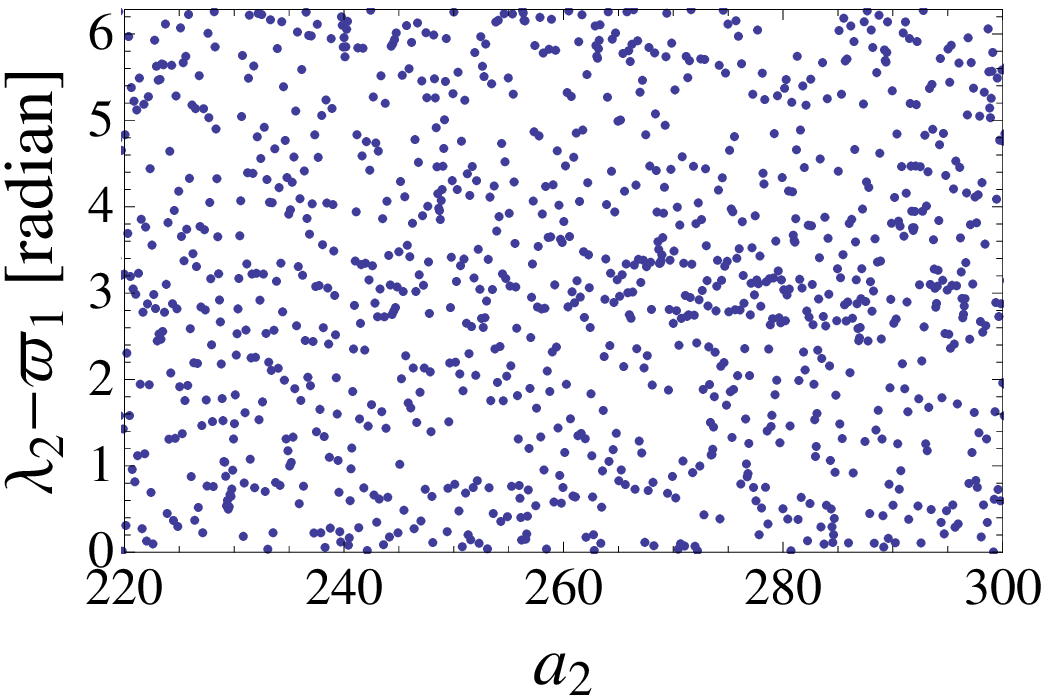}
\includegraphics[width=40mm]{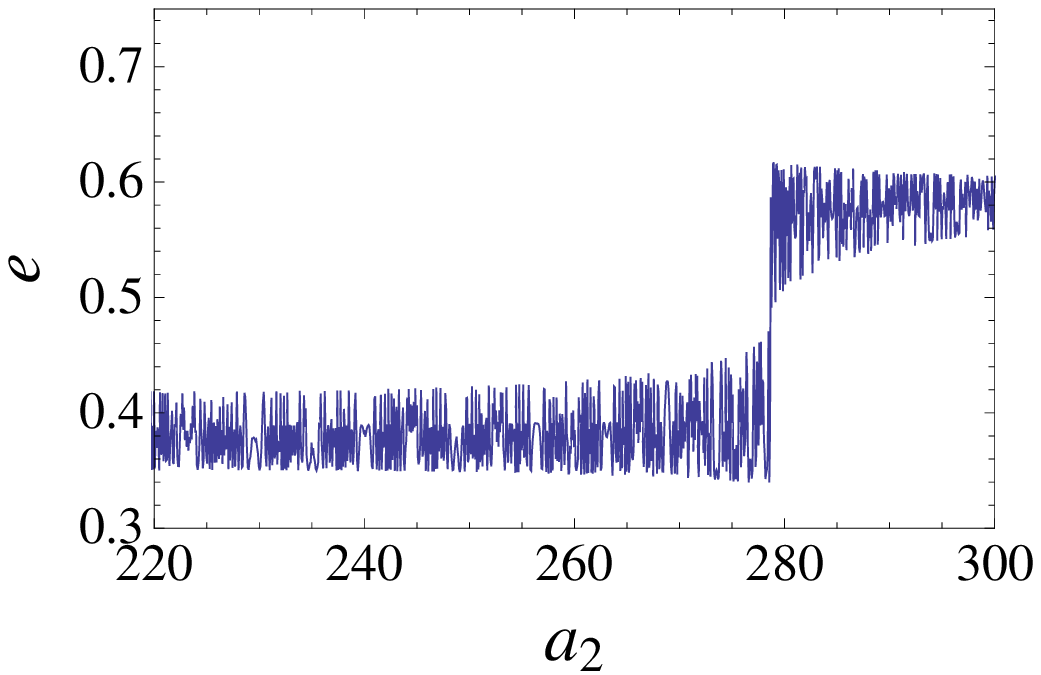}
\includegraphics[width=40mm]{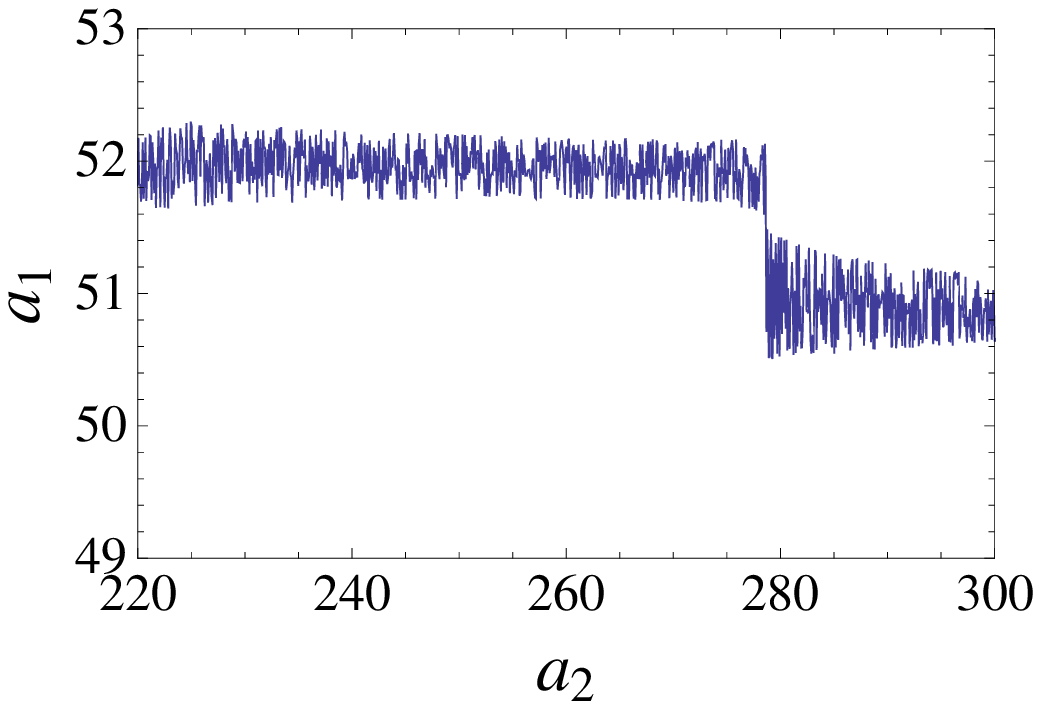}
\caption{ Evolution of a coplanar triple from similar orbital parameters ({\it e.g.} 
$e_{in}\sim0.57$) as Fig.\ref{el1} but with a different relative orbital phase 
(model I). 
The resonant capture is unsuccessful.  We can observe gaps of orbital parameters at the 
resonant encounter around  $a_2\sim 280$.
}
\label{fpot}
\end{figure}

In Fig.\ref{fpot}, the initial inner eccentricity $e_{in}\sim 0.57$ is close to that in 
Fig.\ref{el1}.  But the initial orbital phases are different  between Figs.3 and 4.  
 While evolutions in Figs.3 and 4 are similar down to $a_2\sim 280$,  their subsequent profiles are completely different.
 Around 
the critical epoch  $a_2\sim 280$,  the inner eccentricity $e$  shows a large gap in Fig.\ref{fpot}, but the EMRI is not captured into the resonance, as indicated by the rotating variable 
$\lambda_2-\varpi_1$.  The inner semimajor axis $a_1$ also has a small gap, but the following Tisserand relation (Murray \& Dermott 2000, 
but now for a coplanar system) holds nearly smoothly around $a_2\sim 
280$;
\beq
\frac{a_2}{2a_1}+\sqrt{\frac{a_1}{a_2}(1-e^2)}\simeq const.
\eeq
This relation connects the  gaps for $e$ and $a_1$ in Fig.\ref{fpot}.

For the standard MMRs,  it is well known that the capture becomes 
a stochastic process when we increase  the eccentricity of the perturbed mass \citep{1984CeMec..32..127B,1986sate.conf..159P,1988PhDT........10M,1988Icar...76..295D,2000ssd..book.....M}. 
In addition,  the eccentricity shows a  gap 
if the capture is failed. These interesting characters are successfully 
explained by the Hamiltonian approach. For our unusual resonance, we make  detailed analysis on these issues 
later in \S 6 and 7.

Our main targets in this paper are the EMRI-MBH triple systems in the resonant 
state $\lambda_2-\varpi_1\sim const$.  But it would be 
worth mentioning that  a related resonant structure was identified in the 
ring of Saturn  \citep{1984Icar...60....1P}.
The Titan ringlet has the semimajor axis of
$1.29R_S$ ($R_S\sim 6\times 10^4$km: the radius of  Saturn) and is  in the resonant state  
$
\lambda_{T}-\varpi_{R}\sim const
$
 with  Titan, the largest satellite of Saturn at the distance $\sim 20 R_S$. Here 
 $\varpi_{R}$ is the longitude of the pericenter of the ringlet and 
 $\lambda_{T}$ is the mean anomaly of Titan.
The apsidal precession $\dv_{R}$ of the ringlet is mainly driven by the multiple moments of 
Saturn ({\it e.g.} its quadrupole moment; $J_2=1.6\times 10^{-2}$). This ringlet has a finite 
eccentricity $e_R\sim (2.6\pm 1.4)\times 
10^{-4}$ and a radial width $\sim 20$km. 

{ Another potentially interesting example is the Earth-Moon system. In its 
early history, the  system might be  resonantly 
affected by the Sun as an outer perturber, through the state  
    $\lambda_{Sun}-\varpi_{Moon}\sim const$ \citep{1998AJ....115.1653T,CS}. 
}

\section{relativistic apsidal precession}

As demonstrated in the previous section, our resonant state is characterized by 
the following relation between  the two angular 
parameters $\lambda_2$ and $\varpi_1$
\beq
\eta\equiv \lambda_2-\varpi_1\sim const.
\eeq
Taking the time derivative of this relation, we have
\beq
{n}_2\sim \dv_1 \label{con1}
\eeq
on average (the dot $\dot{}$ representing the time derivative).  Here ${n}_l$ is the angular frequency of the object $m_l$ ($l=1,2$) 
around the central MBH $m_0$, and evaluated with Kepler's third law as
\beqa
n_1&=&\lmk\frac{m_0}{a_1^3}  \rmk^{1/2}\\
n_2&=&\lmk\frac{m_0+m_2}{a_2^3}  
\rmk^{1/2}=m_0^{-1/2}n_1 \lmk  \frac{a_1}{a_2}\rmk^{3/2}\label{n2}
\eeqa
for $m_1\ll m_0+m_2=1$.  To characterize the hierarchy of the inner and outer 
orbits, we introduce the factor $\alpha$ as
\beq
\alpha\equiv\frac{a_1}{a_2}\ll 1.
\eeq
Then the outer frequency is roughly given as
\beq
n_2\sim n_1\alpha^{3/2}
\eeq
for $m_0=O(1)$.

Next we discuss the apsidal precession rate $\dv_1$ of the inner EMRI. As is 
well known for  Mercury, relativistic correction generates the precession 
with the rate 
\beq
{\dot \varpi}_{1r}=\frac{3m_0^{3/2}}{a_1^{5/2}(1-e^2)}=\frac{3pn_1}{1-e^2}\label{pre1}
\eeq
at the 1PN order \citep{1971ctf..book.....L}.  Here, in order to explicitly show the relativistic effects, 
we additionally defined the post-Newtonian parameter $p$ of the EMRI as 
\beq
p\equiv \frac{m_0}{a_1}.
\eeq
In this paper, we only deal with the regime $p\ll 1$ where the PN framework 
works well. The relativistic precession (\ref{pre1}) depends on the eccentricity 
$e$ as $\propto (1-e^2)^{-1}$.  As we see later in \S 5, this dependence becomes 
particularly important for our  unusual resonance.

From Eqs.(\ref{con1})(\ref{n2}) and (\ref{pre1}), we obtain the following relation 
for the onset of the resonance
\beq
\frac{3p}{1-e^2}=m_0^{-1/2}\alpha^{3/2} \label{equi}
\eeq
or equivalently
\beq
a_2=3^{-2/3}a_1^{5/3}(1-e^2)^{2/3}m_0^{-1}. \label{ares}
\eeq
The expression for $e=0$ was studied in the previous paper (Seto 2012, see also 
Hirata 2011) 
and we  
have the relation between the PN parameter $p$ and the orbital hierarchy 
parameter $\alpha$ as $p\sim\alpha^{3/2}/3$. For the eccentric cases shown in 
 the previous section, Eq.(\ref{ares}) provides $a_2\simeq360$ for Fig.\ref{es} and $a_2\simeq285$ for 
Figs.3 and 4,  reasonably reproducing the dependence on the eccentricity $e$.

Eq.(\ref{equi}) is obtained by neglecting influence of the distant outer 
MBH $m_2$ and  assuming that the precession rate $\dv_1$ is dominated by 
the relativistic effect $\dv_{1r}$.  Here we evaluate the Newtonian secular contribution $\dv_{1N}$ due to $m_2$. 
For moderate eccentricity and inclination, the secular effect  $\dv_{1N}$ is 
estimated as \citep{2000ssd..book.....M}
\beq
{\dot \varpi}_{1N}=\frac34\frac{m_2 a_1^{3/2}}{a_2^3 m_0^{1/2}}.
\eeq
Then, at the critical distance (\ref{ares}), we have 
\beq
\frac{{\dot \varpi}_{1N}}{{\dot \varpi}_{1r}}=\frac94 \frac{m_0m_2}{a_1}\ll 1
\eeq
with $p=m_0/a_1\ll 1$ and $m_2<1$. Therefore, the Newtonian  contribution for 
the precession $\dv_1$ would be much smaller than the relativistic one.
{ The distant outer body $m_2$ also has a 1PN effect for the precession $\dv_1$ (see 
the 1PN interaction term in Naoz et al. 2012).  But its magnitude is 
$O(\alpha^{2.5})$ times smaller than Eq.(\ref{pre1}), and not important for the 
precession $\dv_1$. 
  }
We 
hereafter put
\beq
\dv_1=\dv_{1r},
\eeq
as already assumed to derive Eq.(\ref{ares}).

\section{comparison between the first and second order resonances}
The gravitational interaction between the inner and outer orbits of a triple 
system has been perturbatively  analyze with the  disturbing function \citep{2000ssd..book.....M}.
For our resonant state $\eta\sim const$ in a coplanar configuration, the relevant element of the 
disturbing function
is expanded as
\beq
\zeta=e^1 C_{1,0} \cos\eta+e^2C_{2,0} \cos2\eta \label{zeta0},
\eeq
where we take the  terms  up to the order $O(e^2)$. 
 The functions $C_{1,0}$ and $C_{2,0}$ depend on the hierarchy parameter 
$\alpha\ll 1$ of the orbital configuration.  They are explicitly given as
\beq
C_{1,0}=\frac23\alpha-\lmk\frac{\alpha\p_\alpha}2+1\rmk 
b_{1/2}^{(0)}(\alpha)=-\frac{15}{16}\alpha^3 \label{c10}
\eeq
\beq
C_{2,0}=\lmk\frac34+\frac{3\alpha\p_\alpha}4+\frac{3\alpha^2\p_\alpha^2}8\rmk 
b_{1/2}^{(2)}(\alpha)=\frac{15}{8}\alpha^2+\frac{105}{64}\alpha^4 \label{c20}
\eeq
with the Laplace coefficients $b_i^{(j)}(\alpha)$.  In Eq.(\ref{c10}) the first 
term  $2\alpha/3$ is the
indirect part 
 and is canceled by the $O(\alpha)$ term of its direct part.  
As a result, the function $C_{1,0}\propto \alpha^3$ has a stronger dependence on 
the parameter $\alpha(\ll 1)$ 
than the counterpart $C_{2,0}\propto \alpha^2$.  Actually, the second-order one $C_{2,0}$ 
 has the lowest power of $\alpha$ among 
the resonant terms  in the form $\cos N\eta$  with  $N\ge 1$.
 We hereafter neglect the term 
$\propto \alpha^4$ in Eq.(\ref{c20}) and put
\beq
\zeta=-\frac{15}{16} \alpha^3 e \lmk  
\cos\eta-2\frac{e}\alpha\cos2\eta\rmk. \label{zeta}
\eeq
This expression shows that the second order term can dominate the first order 
one even at a small eccentricity $e\sim \alpha$, due to the hierarchy of 
the system $\alpha \ll 1$.

Interestingly, the competition of the two terms can be directly observed as a 
shift of the mean angle $\eta$ of  libration, during the resonant 
amplification of the inner eccentricity $e$.  We now discuss this in some detail.  For 
simplicity, we assume that the dissipative evolution is negligible during one 
libration period.

\begin{figure}
\includegraphics[width=75mm]{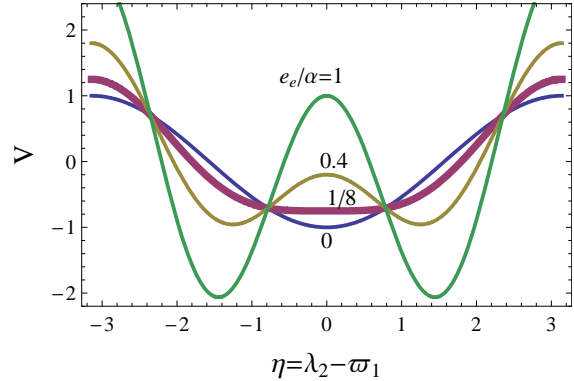}
\caption{ The effective potential $V$ defined in Eq.(\ref{v}). Its shapes are 
plotted for representative values of $e_e/\alpha$. The equilibrium angle $\eta_e$ 
would be the minimum of the potential. For $e_e/\alpha<1/8$, the potential has 
the single minimum at $\eta=0$, dominated by the first order term. In contrast, 
for  $e_e/\alpha>1/8$,  we have two 
minima at $\eta\ne 0$, reflecting the second order component. For larger 
$e_e/\alpha$, the minimal points approach $\pm \pi/2$.
}
\label{fig5}
\end{figure}

First, the system around the resonance can be effectively reduced 
to one dimensional system  (with the  variable $\propto \eta$ and its conjugate 
 momentum $\propto 
 e^2$, see \S 5 for detail). 
The effective Hamiltonian has the resonant term $\propto \zeta$, and the 
variable $\eta$ appears only in this term. Then, from the canonical equation, we 
should have 
\beq
\frac{de^2}{dt}\propto \frac{\p\zeta}{\p \eta}=0
\eeq
at the equilibrium 
point  $(e_e,\eta_e)$.  Thus, for given equilibrium value $e=e_{e}$, 
we 
associate the 
corresponding equilibrium angle  $\eta_e$ as the minimum of the 
following potential $V(\propto \zeta)$
\beq
V\equiv -\cos\eta+\frac{2e_{e}}{\alpha} \cos2\eta.\label{v}
\eeq
The shape of this potential is shown in  Fig.\ref{fig5} for representative 
values of the ratio
$e_{e}/\alpha$. 
The positions of the potential minima qualitatively change at the critical 
value $e_e=\alpha/8$.  In Table 1, we present its value for models I-III.
At  $e_e<\alpha/8$, the potential $V$ is dominated by the first order 
term and we have the equilibrium angle
\beq
\eta_e=0,~~~~~(e_{e}<\alpha/8). \label{th1}
\eeq
When increasing $e_{e}$ beyond the critical value $\alpha/8$, the angle 
$\eta_e$ starts to move as 
\beq
\eta_e=\pm \arccos(\alpha /8e_e),~~~~~(e_e>\alpha/8).\label{th2}
\eeq
We have $\eta_e\simeq \pm \pi/2$ for $e_e\gg \alpha/8$, dominated by the second 
order term in Eq.(\ref{v}).

\begin{figure}
\includegraphics[width=40mm]{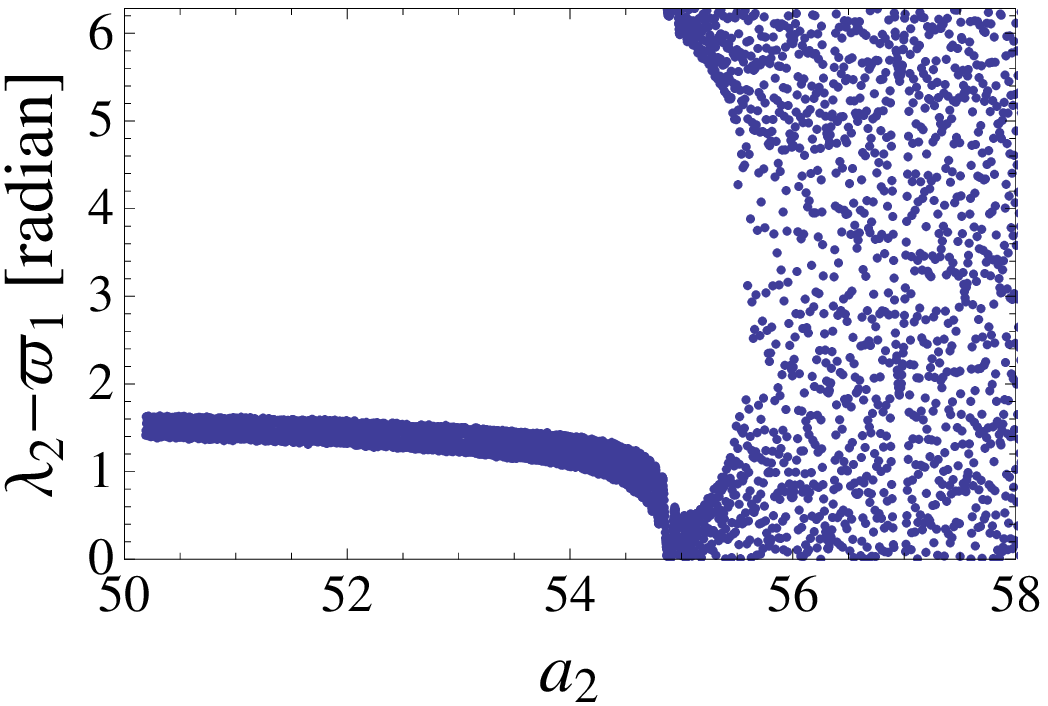}
\includegraphics[width=40mm]{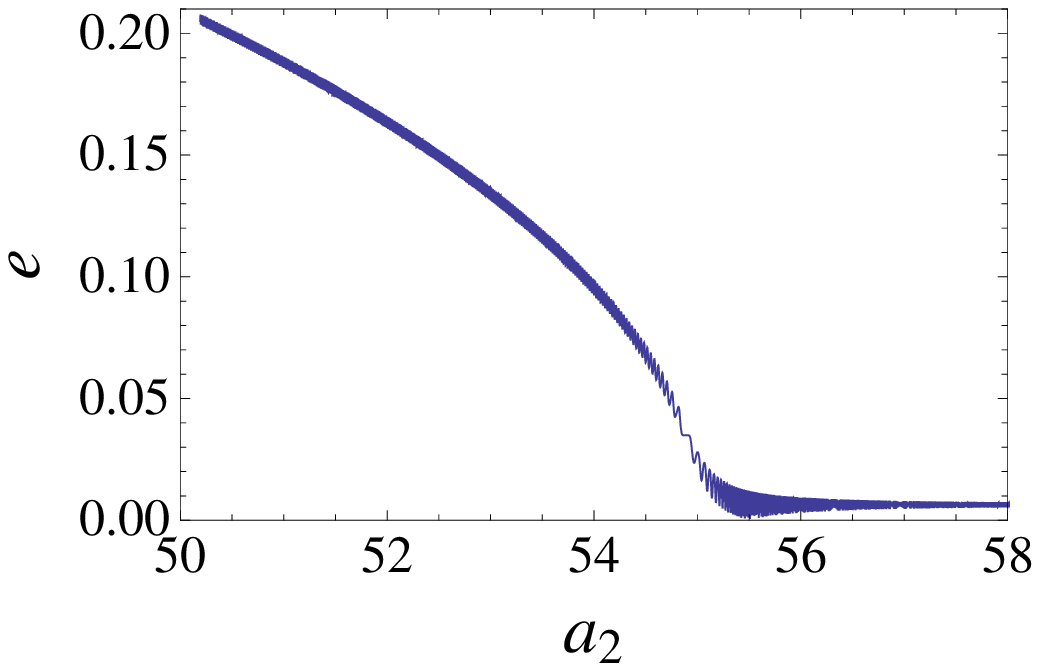}
\includegraphics[width=40mm]{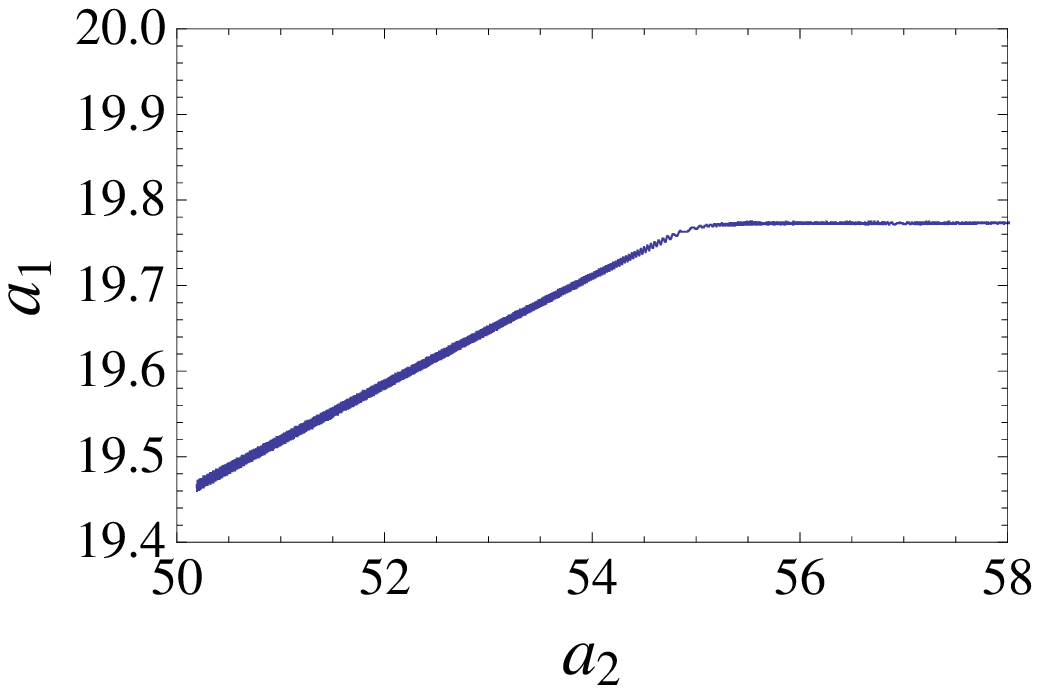}
\caption{ Evolution of the orbital elements of the coplanar inner EMRI for model III.  Due to the small outer mass $m_2$,  the forced components can be 
suppressed, compared with cases I and II.  We have a small initial inner
eccentricity    $e\sim0.0065$.  The EMRI is captured by the outer MBH binary around 
$a_2\sim 55$. We can clearly observe the shift of the equilibrium angle $\eta_e$. 
}
\label{case3}
\end{figure}

Now we examine our simple model (\ref{th1}) and (\ref{th2}) for the equilibrium 
resonant angle, by using numerical simulations.
In Fig.\ref{case3}, we show the evolution of orbital parameters for model III.  
Owing to its small outer mass $m_2$, the forced 
eccentricity  is small at the early stage, and this model allows us to make a suitable demonstration for the present analysis. 
 
\begin{figure}
  \includegraphics[width=70mm]{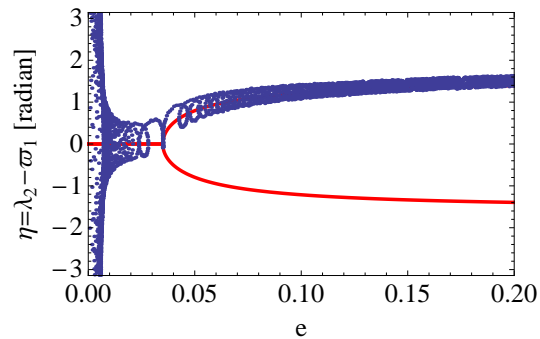} 
  \caption{   Correspondence between the inner 
eccentricity $e$ and the resonant angle $\eta=\lambda_2-\varpi_1$.  The points 
are obtained from  the run  
shown in Fig.6.   The red curves are the analytical predictions
(\ref{th1}) and (\ref{th2}) for the equilibrium points $(e_e,\eta_e)$ with the transition 
eccentricity at  $\alpha/8=0.035$.    
 }
\label{case32}
\end{figure}

The EMRI is resonantly captured by the outer MBH binary around $a_2\sim 55$, and its 
eccentricity $e$ starts to increase afterward. Here the critical eccentricity for 
the onset of the shift of the equilibrium angle is $\alpha/8\sim 0.035$.  Since the libration can be 
regarded as a circulation around the equilibrium point  $(e_e,\eta_e)$, the mean 
value of the  
libration would be close to the  equilibrium point  $(e_e,\eta_e)$,  at least for a small libration amplitude.
To directly show the shift of the angle $\eta_e$ thorough the  
resonant amplification  of the inner eccentricity $e$, we plot the combination 
$(e,\eta)$ in Fig.\ref{case32} together with the analytical model (\ref{th1}) and 
(\ref{th2}).

Fig.\ref{case32} shows that, 
even though the dissipative time scale is not sufficiently long compared with 
the libration period,  the simple analytical prediction shows a good agreement 
with the numerical one.  For a larger libration amplitude as in Fig.\ref{el1},  the 
potential wall of $V$ around $\eta=0$ (see Fig.\ref{fig5}) is easily crossed over, and 
the angle $\eta$ moves around  a broad region, leaving a small excluded regime 
near  $\eta=\pi$.

Note that also in Fig.\ref{es},  the  angle $\eta$ finally localizes around $\pi/2$, as 
in the case of Fig.6.  But this should be regarded as a mere coincidence.  Later 
in  \S 6 and 7, we deal with a large sample of numerical simulations for models 
I and II. 
Among them, 
there are no definite asymmetries for the preference of the two potential 
minima at $\eta=\pi/2$ and $-\pi/2$ (equivalently $3\pi/2$).

In summary, due to the hierarchy of the orbits with $\alpha\ll 1$, the second 
order term $\propto e^2\alpha^2\cos2\eta$ could become more important than 
the first order one $\propto e \alpha^3\cos\eta$, even for a small 
eccentricity $e\sim \alpha$.  We can observe the resultant shift of the mean 
(equilibrium) angle $\eta_e$ during the resonant amplification of the inner 
eccentricity $e$.

\section{Hamiltonian Approach}

In this section, we apply the Hamiltonian approach for the resonant dynamics of 
the EMRI-MBH triple systems with $\lambda_2-\varpi_1\sim const$. By taking appropriate set of conjugate variables, 
the dynamics around the resonant encounter can  be reduced to a simple one 
dimensional system 
\citep{1972MNRAS.160..169S,1979CeMec..19....3Y,1982CeMec..27....3H,1983CeMec..30..197H,1986sate.conf..159P,2000ssd..book.....M}.  For  the standard MMRs such as 
2:1 or 3:1 resonances, 
this approach successfully explains characteristic phenomenon around the 
resonant  
encounters \citep{1984CeMec..32..127B,1986sate.conf..159P,2000ssd..book.....M}. Our 
aim here is  to extend it for our unusual resonance. As  detailed 
descriptions of the approach for the standard MMRs can be found in the 
literature and many of them are shared with our resonance,  it would be unfruitful to lengthily 
expound  all the involved steps.  We rather follow the comprehensive 
formulation given in \S 8.8 of \cite{2000ssd..book.....M}  and explain the modifications necessary for 
our specific resonance.

\subsection{Simplified Hamiltonian for Second Order Resonance}
Based on the results in the previous section, we analyze the second order 
resonance  with the resonant variable
\beq
\theta_1=j\lambda_2+(2-j)\lambda_1-2\varpi_1 \label{varia},
\eeq
identical to Eq.(8.78) of MD with $k=2$. While we are mainly interested in 
the specific case $j=2$, 
  we do not fix the parameter $j$ at this stage, in order to enable a simple 
  comparison with the standard second-order MMRs corresponding to $j>2$.

As explained in MD, the variable $\theta_1$ has the conjugate momentum 
$\Theta_1$ defined by 
\beqa
\Theta_1&=&\frac{m_1}2 \sqrt{m_0a_1}\lmk 1-\sqrt{1-e^2}  \rmk\\
&\simeq&\frac{m_1}4 
\sqrt{m_0a_1}\lmk e^2 +\frac{e^4}4+\cdots  \rmk. \label{Th}
\eeqa
We should notice that this momentum is directly related to the inner 
eccentricity $e$ as $\Theta_1\propto e^2$.  In our analytical studies below, we 
make perturbative expansions,  assuming $e\ll 1$.

Among multiple terms in the Hamiltonian (8.98) of  MD (denoted as $\cal H$), the key element for our 
unusual resonance is the 
following one
\beq
-k \Theta_1 {\dot \varpi}_{sec}\equiv X \label{nx}
\eeq
with $k=2$ for the present analysis. Here, the notation $\dv_{sec}$ in MD
represents  the 
secular  precession rate of the inner pericenter and is identical to the relativistic
apsidal precession $\dv_{1r}$ under our prescription in \S 3 (hereafter 
using $\dv_{1r}$ in stead of $\dv_{sec}$).

With respect to the canonical equation
\beq
\frac{d\theta_1}{dt}=\frac{\p {\cal H}}{\p \Theta_1},
\eeq 
the term $X$ in the total Hamiltonian $\cal H$ has a role to provide the secular 
contribution  $-k \dv_{1r}$ for the time derivative $d\theta_1/dt$.  Therefore, 
we should have  the  equation below
\beq
\frac{\p X}{\p \Theta_1}=-k {\dot \varpi}_{1r}.\label{dxdt}
\eeq
Meanwhile, as given in Eq.(\ref{pre1}), the relativistic precession rate $\dv_{1r}$ has 
the following form at 1PN order
\beq
{\dot \varpi}_{1r}=\frac{3pn_1}{1-e^2}\simeq 3pn_1(1+e^2),
\eeq
and the rate ${\dot \varpi}_{1r}$ itself depends on the conjugate momentum $\Theta_1\propto 
e^2$.  Thus we have the following perturbative solution $X$ for Eq.(\ref{dxdt})
\beqa
X&=&-3kp  n_1 \Theta_1 \lmk 1+ \frac{ e^2}{2} \rmk\\
&=&-3kp  n_1 \Theta_1 \lmk 1+ \frac{ 2\Theta_1}{m_1\sqrt{m_0a_1 }}  \rmk \label{solx}
\eeqa
expanded in terms of the momentum $\Theta$, instead of the eccentricity $e$.
Note that this solution is  different from the naive expression (\ref{nx}) that 
is perturbatively expanded as
\beq
-k \Theta_1 {\dot \varpi}_{sec}=-3kp  n_1 \Theta_1 \lmk 1+ \frac{ 4\Theta_1}{m_1\sqrt{m_0a_1 }}  \rmk .
\eeq 
The quadratic term 
$\propto \Theta_1^2$ plays a critical role for our resonance, as we see in the 
next subsection. This term originates from the dependence $\dv_{1r}\propto (1-e^2)^{-1}$.

One might has an impression that  the present derivation for Eq.(\ref{solx}) 
is phenomenological, as it is constructed to reproduce the desired precession 
rate $\dv_{1r}$.   But we can actually  derive the  term 
(proportional to $e^2+3e^4/4+\cdots$) identical to $X$ in Eq.(\ref{solx}), starting 
directly from the 1PN Hamiltonian $H_1$ in Eq.(\ref{ham3}) (see Eq.(28) in Naoz 
et al. 2012).  We took the above route to elucidate the modification relative to the 
 typical analysis for the standard MMRs.

With the explicit form of the relativistic correction $X$ in hand, we can next apply
the standard arguments in MD to derive a simplified Hamiltonian for MMRs. After some calculations ({\it e.g.} introducing 
the new conjugate variables $\theta_1'=\theta_1/2$ and $\Gamma\equiv 2\Theta_1$), 
 we have the following Hamiltonian  (corresponding to Eq.(8.102) of MD) 
\beq
{\cal H}^\dagger={\bar \alpha}\Gamma+{\bar \beta}\Gamma^2+2{\bar \epsilon}\Gamma 
\cos2\theta_1'. \label{oh}
\eeq
Here the coefficients $\ba, \bb$ and $\be$ are given as 
\beq
{\bar \alpha}=\frac{(j-2)n_1-jn_2+6p n_1}2, \label{ba}
\eeq
\beq
{\bar \beta}=\frac3{8}\lmk \frac{(j-2)^2}{m_1 a_1^2} +\frac{j^2}{m_2 a_2^2}\rmk +3 \frac{p}{m_1 a_1^2},\label{bb}
\eeq
\beq
\bar \epsilon=C_{j,j-2} n_1 \frac{ m_2}{m_0}\alpha \label{c24}.
\eeq
In Eqs.(\ref{ba}) and (\ref{bb}),  the terms proportional to the PN parameter $p$ clearly show the relativistic 
corrections.  The factor $C_{j,j-2}=C_{2,0}$ for $j=2$  was already given in 
Eq.(\ref{c20}).  In the right-hand side of Eq.(\ref{bb}),  the first parenthesis 
appears in the standard MMRs and has its origin in the Keplarian terms in the 
triple system (see MD). Its second term ($\propto p$) is due to the quadratic term $\propto 
\Theta_1^2$ in the secular correction $X$ for the relativistic apsidal precession.

We further make  transformation of variables as follows
\beq
\Phi=\frac{\Gamma \bb}{2\be}=\frac{\Theta_1 \bb}{\be},~~\tau={2\be}{t} \label{Ph}
\eeq
\beq
\phi=\cases{
\theta_1' & (${\bar \epsilon}<0$) \cr
\theta_1'+\pi  & (${\bar \epsilon}>0$) \cr
},\label{phi}
\eeq
and finally obtain the rescaled Hamiltonian
\beq
H=\Phi^2 +\bd~ \Phi+\Phi \cos2\phi \label{sh}
\eeq
with the single parameter $\bd$ defined by
\beq
\bd=\frac{\bar \alpha}{2\bar \epsilon}.
\eeq
The associated canonical equations are written as
\beq
\frac{d\Phi}{d\tau}=-\frac{\p H}{\p \phi},~~\frac{d\phi}{d\tau}=\frac{\p H}{\p 
\Phi}. \label{ce}
\eeq

The rescaled Hamiltonian (\ref{sh}) is slightly different from the related 
expression (8.116) in MD, but identical to those in \cite{2006MNRAS.365.1367Q} and \cite{2011MNRAS.413..554M}.  We adopt the present form, in order to use these two references 
later and   discuss whether  evolution of the parameter $\bd$
 can be regarded as adiabatic for our resonant dynamics.

Roughly speaking, this parameter $\bd$ represents an effective distance to the 
resonance.  Due to the GW emission, the orbits of the EMRI-MBH triple system decay 
gradually, and the parameter $\bd$  varies accordingly.

We now estimate the transition rate $d\bd/d\tau$.
First, apart shortly from the triple systems, we consider a simple binary with a semimajor axis $a$, an eccentricity $e$ and 
masses  
$m, m'$.   Its orbital decay rate $da/dt$  by GW 
emission is given as \citep{1964PhRv..136.1224P}
\beq
\frac{da}{dt}=-\frac{64}5\frac{mm'(m+m')}{a^3(1-e^2)^{7/2}}\lmk1+\frac{73}{24}e^2+\frac{37}{96}e^4  \rmk.\label{od}
\eeq

Next, for our triples,  we assume that,  before the  resonant encounters, the EMRI 
and MBH binary independently evolve with Eq.(\ref{od}).  Then we obtain
\beqa
\frac{d\bar \alpha}{dt}&=&\frac{48}5 \Big[m_1 m_0^{5/2} a_1^{-13/2}\frac{\lnk (j-2)a_1+ 
10m_0\rnk}{(1-e^2)^{7/2}}\nonumber \\
& &\times \lmk1+\frac{73}{24}e^2+\frac{37}{96}e^4  \rmk
 -j m_0 m_2 
a_2^{-11/2}   \Big] \label{da}
\eeqa
for $e_2=0$.
With the scaled time $\tau$,  we can obtain the transition rate as 
\beq
\frac{d\bd}{d\tau}=\frac1{2\bar\epsilon}\frac{d\bar\alpha}{dt} 
\frac{dt}{d\tau}=\frac{1}{4\bar \epsilon^2}\frac{d\bar\alpha}{dt}.\label{ddd}
\eeq

For a given EMRI-MBH triple around the resonant encounter, we can now analyze its 
evolution through the one-dimensional  rescaled Hamiltonian 
(\ref{sh}). The information of the original triple system is converted to (i) the new 
variables $(\phi,\Phi)$, (ii) the parameter $\bd$ and (iii) its time derivative 
$d\bd/d\tau$.  In practice,  this mapping can be  made with Eqs.(\ref{ba})-(\ref{phi}) and (\ref{da})-(\ref{ddd}). In the 
next subsection, we concretely study the relation in the test particle limit 
$m_1\to 0$. But, here, we derive a result valid also for $m_1\ne 0$.

To realize a capture ({\it i.e.} transition of $\phi$ from rotation to libration) with Eq.(\ref{ce}), the resonance should be crossed in the direction 
$d\bd/d\tau<0$ (Peal 1986; MD).  In the cases of standard MMRs, this 
corresponds to relatively approaching orbits.
For example, to be captured into 
the 3:2 resonance,  the ratio of the orbital periods should change in the 
direction of $1.6\to 1.5$ not 
$1.4\to 1.5$. With Eqs.(\ref{ares}) and (\ref{da}) for $j=2$, the inequality $d\bd/d\tau<0$ is rewritten as
\beq
\frac{a_1}{a_2}>\sqrt{\frac53 (1-e^2)^{-5/2}\lmk1+\frac{73}{24}e^2+\frac{37}{96}e^4  
\rmk \lmk \frac{m_0 m_1}{m_2}\rmk^{1/2}} \label{cri}.
\eeq
For the specific case $e=0$, this expression agrees with that derived and 
examined  in 
Seto  
(2012).  Note that our labels $(0,1,2)$ for the three masses are different from 
those in Seto (2012). 

\subsection{Test Particle Limit}
Here we discuss the mapping between the EMRI-MBH triple system and the simplified 
Hamiltonian system (\ref{sh}) in the test particle limit $m_1\to 0$. In this 
limit, we can easily control the relative orbital evolutions of the triple 
system in numerical simulations, and, furthermore, the role of the 
post-Newtonian corrections  becomes transparent.

From Eqs.(\ref{Th})(\ref{bb})(\ref{c24}) and (\ref{Ph}), the inner eccentricity 
$e$ is related to  the momentum $\Phi$ as
\beq
\Phi=\frac{m_0 a_1^2}{20m_2 \alpha^3} \lkk \frac{(j-2)^2}{a_1^2}+\frac{m_1j^2}{m_2a_2^2}+\frac{8p}{a_1^2}  
\rkk e^2 \label{Ph2}.
\eeq
In this relation, we pay our attention to the dependence of the mass parameter 
$m_1$. We can put $p=0$ in the traditional analysis of the standard MMRs with $j>2$ (see Eq.(8.109) of MD). However, for our unusual one 
with $j=2$,  the mapping (\ref{Ph2}) becomes singular $\Phi\propto 
m_1e^2$ in the limit $m_1\to 0$, if the relativistic effect is dropped 
with $p=0$.  Therefore, interestingly, the regularity of the mapping (\ref{Ph2}) 
is maintained by the post-Newtonian correction ($\propto p$) for our resonance with 
$j=2$ as
\beq
\Phi= \frac{2pm_0 a_2^3e^2}{5m_2 a_1^3}
\eeq
without depending on $m_1$.  As mentioned earlier, the  post-Newtonian term in Eq.(\ref{Ph2})
comes from the quadratic term of the momentum $\Phi$ in Eq.(\ref{solx}) and 
intrinsically from the dependence of the precession rate on the eccentricity as 
shown in Eq.(\ref{pre1}).

Now we derive formulae specifically for $j=2$ with the rescaled Hamiltonian
\beq
H=\Phi^2 +\bd~ \Phi+\Phi \cos2\phi \label{sh2}.
\eeq
 The variable $\phi$ is related 
to the original resonant angle $\eta\equiv \lambda_2-\varpi_2$ as
\beq
\phi=\eta+\pi.
\eeq
After some algebra with Eq.(\ref{ares}),  the principal quantities for the rescaled Hamiltonian are 
given by the original parameters as
\beq
\Phi=De^2
\eeq

\beq
D\equiv \frac{2a_1}{45 m_0 m_2 } \label{cd}
\eeq

\beq
\frac{d\bd}{d\tau}=-\frac{512m_0^{3/2}}{3^{1/3 }375  m_2 a_1^{13/6}}.\label{dd}
\eeq
Note that the transit speed $d\bd/d\tau$ is nearly constant  around the resonant 
encounter, and we omit the expression for the time dependent parameter $\bd$ itself.

In Table.1, we provide the transit speed $d\bd/d\tau$ as well as the coefficient 
$D$.
The former is an useful measure to discuss the adiabaticity of the time evolution of the 
parameter $\bd$ at the resonant encounter.

For comparison, including only the first order 
resonant term $\propto e$ in Eq.(\ref{zeta}), we  derive the relevant expressions for the test particle 
limit.
In Appendix A, we summarize the results.  Again, we have a regular 
mapping between the momentum $\Phi$ and the inner eccentricity $e$, due to the 
PN correction.

For a Newtonian apsidal precession induced by multiple moments of masses, the 
precession rate  at $e\ll 1$ generally has correction for the eccentricity $e$ starting from 
$e^2$ \citep{1939MNRAS..99..451S}.  
Therefore,  the mapping to a corresponding rescaled Hamiltonian becomes regular,  as for the 
relativistic one discussed above. For example, the precession rate of a test 
particle  due to the quadrupole moment $J_2$ of the 
central body  is expanded as $\dv\propto J_2 (1+2e^2+\cdots)$.   

In the next three sections, using  the mapping from the 
EMRI-MBH triple system, we make  quantitative 
predictions  on the resonant dynamics and compare them with numerical simulations.
Below, we limit our analysis to the test particle limit $m_1\to 0$.

\section{capture rate}

In this section, we study whether the analytical model based on the rescaled Hamiltonian can reproduce the capture 
rate estimated from numerical simulations. 

For this comparison, we obtained the capture rate from the numerical side in the following manner.  First,  for models I and II, we took various ($\sim 13$) inner
initial eccentricities $e_{in}$ between $\sim0$ and $\sim 0.7$.  For each 
eccentricity, we assigned  an initial outer radius $a_{2in}$ larger than 
Eq.(\ref{ares})  
 to assure a resonant encounter, and made 20 runs starting from randomly distributed relative 
orbital phases. 
Therefore, the total number of the runs is $\sim 2\times 13\times 20\sim 500$.  We 
judge a run as  a resonantly captured event when the angle $\eta=\lambda_2-\varpi_1$  
has a single excluded range $\Delta \eta$ larger than $\pi/10$ for minimum duration
$a_2/10$ (and also $a_2/20$ for comparison) in terms of the decaying  outer radius $a_{2}$ (see {\it e.g.} 
Figs.2 and 3).  
{  
  An MMR is often identified with a sharp concentration of a 
resonance angle, as found in Fig.\ref{es}. But, here, we  employed the criteria 
for $\Delta \eta$ 
 to handle  the resonant state with  
large libration amplitudes, 
as demonstrated in Fig.\ref{el1}. }

Then, for each initial eccentricity $e_{in}$, we counted the number of 
captured  
events among the associated 20 runs and { roughly} obtained the capture rate.  We provide the numerical results 
in Fig.8 with the blue circles { for minimum duration $a_2/10$.  Up to 
moderate eccentricity, we obtain the same results   for the minimum duration 
 $a_2/20$. 
Only when they are different,  we added the latter with the open squares. }. 
 As shown in Figs.2-4,  the eccentricities of the EMRIs 
are always  oscillating to some extent. To handle this,  we  took  time averaged eccentricity for each run 
at its early stage, and subsequently evaluated the mean value among the 20 runs. The initial
eccentricities in Fig.8 are made up in this way.

\begin{figure}\label{ratec}
  \includegraphics[width=70mm]{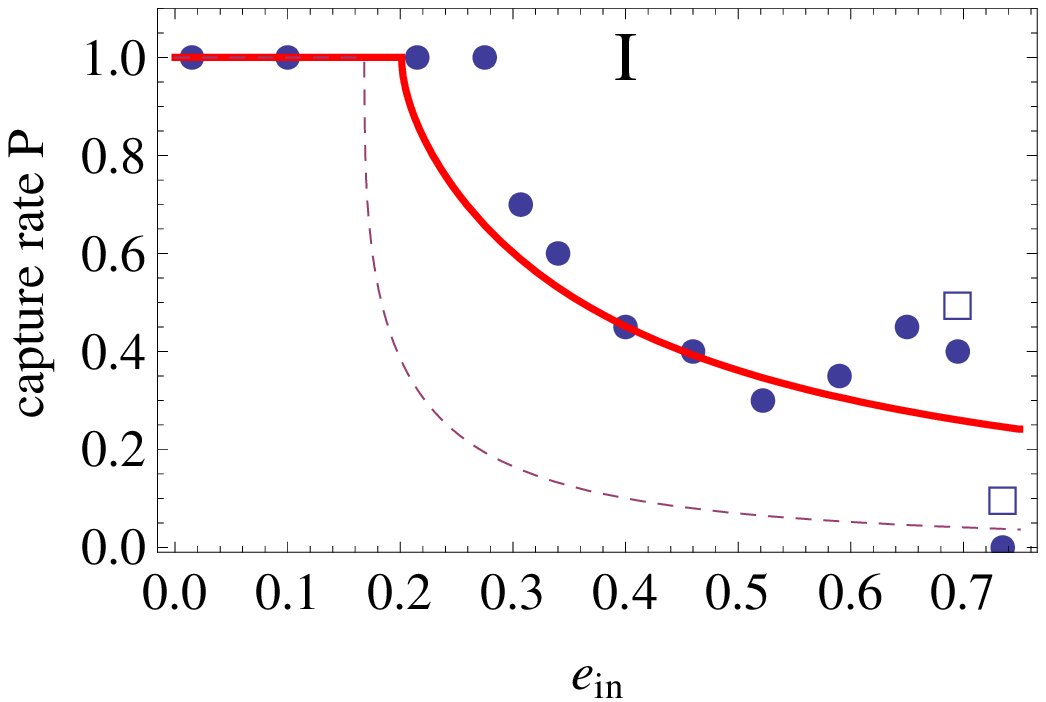} 
  \includegraphics[width=70mm]{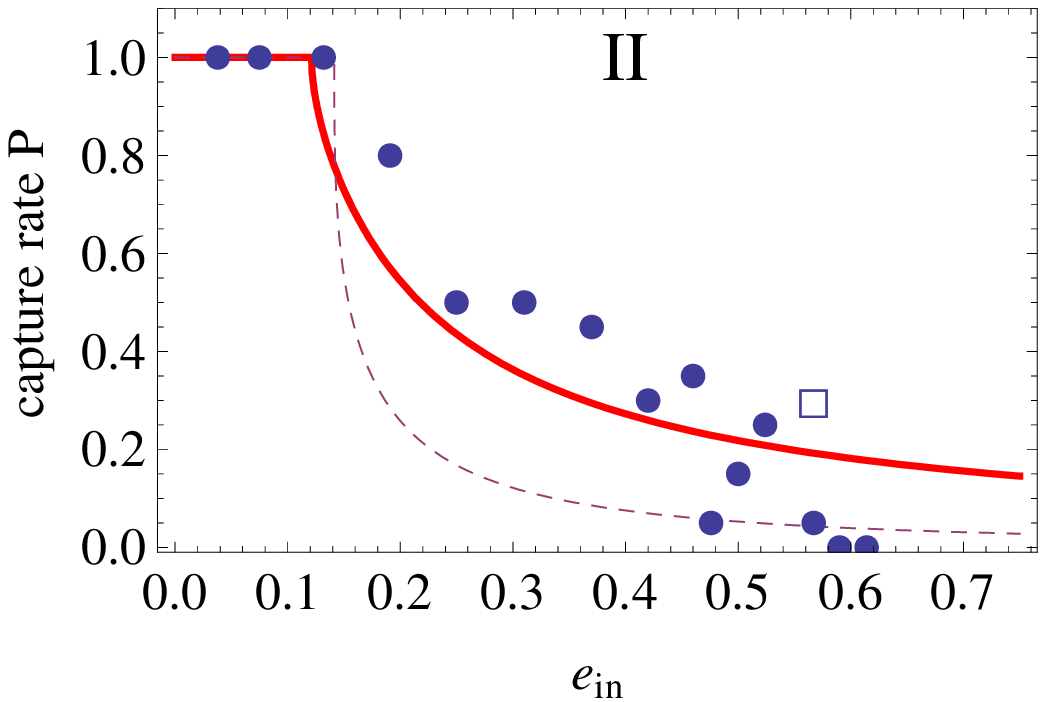}
  \caption{Resonant capture rate of an inner EMRI by a coplanar outer MBH.   
The results are given as  functions of the inner eccentricity $e_{in}$  
for models I and II. The 
solid curves are the analytical predictions $P_2$ for 
the second order term $\propto \cos2\eta$ and the dashed curves are $P_1$  for the 
first-order term $\propto \cos\eta$.  { The circles are numerical results for 
minimum duration $a_2/10$, and the additional open squares are for $a_2/20$ 
(shown only if different).} Each of 
them is estimated from 20 runs.
 }
\end{figure}

\begin{figure}\label{gapc}
  \includegraphics[width=70mm]{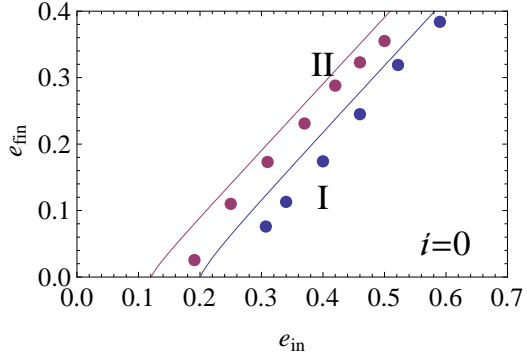} 
  \caption{ Gaps of the eccentricities $e$ of inner EMRIs with failed captures (as shown in Fig.\ref{fpot}).  With the solid curves, 
we plot the analytic
correspondences between the gapped eccentricities before ($e_{in}$) and after ($e_{fin}$) 
the resonant encounter for models I and II. 
The circles are 
the mean eccentricities of the numerical runs accompanied by the  gaps.
 }
\end{figure}

Next we  analytically estimate the capture rate  through the rescaled Hamiltonian 
(\ref{sh2}),  
following Borderies \& Goldreich (1984).  In Appendix B, we briefly describe 
their results.  As we have already discussed,  the second order 
resonance is relevant for our systems (unless $e_{in}$ is less than $O(10^{-2})$, see Table 1).   Therefore, we mainly use their results 
for $k=2$ (see \S B1).

For an initial eccentricity $e_{in}$, 
the analytical  rate $P_2$ is obtained 
through the projected initial momentum 
\beq
\Phi_{in}=D e_{in}^2
\eeq
 for the rescaled Hamiltonian. 
There is  a critical value $\Phi_{cr}=1$, corresponding to $e_{in}=\sqrt{\Phi_{cr}/D}=0.20$ (model 
I) and 0.12 (model II) that are much larger than $\alpha/8=O(10^{-2})$ (see Table 1).   The capture occurs at 
100\% for $\Phi_{in}<\Phi_{cr}$.  Meanwhile, for  $\Phi_{in}>\Phi_{cr}$, the 
capture becomes a stochastic process with the rate $P_{2}$ defined implicitly 
by Eqs.(\ref{p2}) and (\ref{p2d}).

In Fig.8,  with the solid curves, we present the analytical rates $P_2$.  At 
$e_{in}\lsim 0.4$, they show reasonable agreements with the numerical ones. But, 
at larger eccentricities $e_{in}\gsim 0.5$,  we have significant discrepancies. 
This is not surprising, since we  made, at least,  various approximations, valid 
only for
$e_{in}^2 \ll 1$.  For reference, we also show the rate $P_{1}$ expected for 
the fist order resonance (see \S B.2), but it poorly fits the numerical data, as 
expected. { Note also that, for higher eccentricities, the numerical results 
 are affected by the applied  conditions for identification of the resonances.}

As we explained earlier, the critical eccentricity $\sqrt{\Phi_{cr}/D}$ 
characterises the overall shape of the capture rate.  Here we should comment on  
 its magnitude for the standard second-order MMRs (with the variable (\ref{varia}) 
 at $j>2$).  For these resonances, dynamical stability of orbits requires 
 $m_2\ll m_0$ (assuming $m_1<m_2$).  Then we can show a scaling behaviour 
 $D=\Phi/e^2\propto 1/m_2$ and obtain the critical eccentricity 
 $\sqrt{\Phi_{cr}/D}\ll 1$ much smaller than our hierarchical one with 
 $m_0/m_2=O(1)$ (as in model I).  Therefore, for the standard second-order MMRs, 
 perturbative expansion of the eccentricity is more effective in the regime where the resonant capture is probable ({\it e.g.} $P_2>0.1$), unlike our hierarchical one with larger $\sqrt{\Phi_{cr}/D}$.
We can make similar arguments for the first-order resonances.

For the analytical predictions in Appendix B,  we fully use the arguments based on 
 the 
adiabatic invariant that is conserved for a  transit speed $|d\bd/d\tau|$ much smaller 
than the libration frequency \citep{1969mech.book.....L}.  To examine the impacts of finiteness of 
$|d\bd/d\tau|$  on the resonant dynamics,  Quillen (2006) and Mustill \& Wyatt 
(2011) numerically studied dependence of the capture rate on the transition speed 
$|d\bd/d\tau|$.  Their results (see {\it e.g.} Fig.2 in Mustill \& Wyatt 2011) indicate that 
the  
adiabatic approximation would be efficient for $|d\bd/d\tau|<0.1$.  As 
shown in Table.1,  two models I and II well satisfy this criteria. 
For a coplanar EMRI-MBH 
triple of comparable MBHs ($m_2/m_0=O(1)$) with converging orbits $d\bd/d\tau<0$, we generally have $|d\bd/d\tau|\ll 0.1$, unless the target EMRI is highly 
 relativistic.

\section{gap of inner eccentricity}
It is well known that, for the standard MMRs, the eccentricity 
of  
a perturbed body shows a gap when the resonance is encountered but capture 
results in failure ({\it e.g.} Peale 1986; MD; see also Amaro-Seoane et al. 2012). This phenomena is well explained by the Hamiltonian approach in the associated phase space, 
as a rapid change of rotational motion at the separatrix crossing.  In Fig.\ref{fpot},  we can 
observe a similar gap of the eccentricity for the run without a resonant capture.

In 
order to further examine validity of  our Hamiltonian model extended for the 
unusual resonance,  we analyze the correspondence of the two eccentricities;  $e_{in}$ 
 (before the encounter) and $e_{fin}$ (after the encounter).  We derive  an analytical prediction using the rescaled 
 second-order Hamiltonian (\ref{sh2}) and compare it with the 
 numerical simulations done in \S 6. 

We first discuss the analytical approach in which the correspondence between the two
eccentricities is equivalent to the relation between the initial momentum 
$\Phi_{in}=De_{in}^2$ and the final one $\Phi_{fin}=De_{fin}^2$ both  far from 
the resonant encounter.  For a given initial momentum
$\Phi_{in}$, the parameter $\bd$  at the separatrix crossing is given in a 
somewhat complicated form as in Eqs.(\ref{p2d}) and (\ref{kai}).  But, because of a simple 
expression for a define integral, we have the following concise relation between 
 the two momenta \citep{1988PhDT........10M}
\beq
\Phi_{in}+\Phi_{fin}=-\bd. \label{plus}
\eeq
Note that the separatrix relevant for our analysis is formed at $\bd <-1$ where 
the capture rate becomes less than unity.  In Eq.(\ref{plus}),  we have 
$\Phi_{in}=1$ and $\Phi_{fin}=0$ for $\bd=-1$.

We can now obtain the desired correspondence $\Phi_{in}\to \Phi_{fin}$ 
(equivalently $e_{in}\to e_{fin}$) through the intervening parameter $\bd 
(<-1)$. 
In Fig.9, we show this analytical correspondence for  models I and II with the 
solid curves.

We also analyze the  samples of the numerical simulations described 
in the previous section.  In Fig.9,  the numerical results are presented with 
the filled circles. 
In numerical data,  formation of a  gap can be easily identified as a instantaneous event, compared with 
  continuation of  a resonant state.
We can observe small systematic deviations between the analytical 
and numerical results.  But, as a whole,  the simple analytical predictions show 
reasonable agreements with the numerical ones that were obtained after rather complicated 
dynamical evolutions.

\section{inclined orbits}
So far we have discussed the resonant dynamics for coplanar orbits. In this 
section, we extend our study to inclined orbits.  We use the parameter $i$ as the relative 
inclination angle between the inner and outer orbits.

{  In Fig.10 we provide a numerical example for inclined orbital 
configurations. This triple 
 system has a  small initial inclination angle 
$i=5.8^\circ$,  but its initial  semimajor axes $a_1$, $a_2$ and  
eccentricity $e$ are close to those in  Fig.2. We find that the overall 
evolution  
of the three quantities $\lambda_2-\varpi_1$, $e$ and  $a_1$ are similar to Fig.2.

Note that the inclination angle $i$ 
stays nearly at a constant value.  This is in accord with the simplified Hamiltonian 
approach that has only two dynamically important variables $e$ and 
$\lambda_2-\varpi_1$ in the present eccentricity resonance. 

In Fig.11, we show the results from a larger inclination angle $i=60^\circ$. 
In the lower right panel,  evolution of the inclination angle $i$ is 
presented in a geometric  form $\cos i$.
We can 
observe oscillation of  $\cos i$.
 But, around the resonant encounter $a_2\sim 305$, its amplitude is much smaller than that of the eccentricity $e$.   

}

\begin{figure}
\includegraphics[width=40mm]{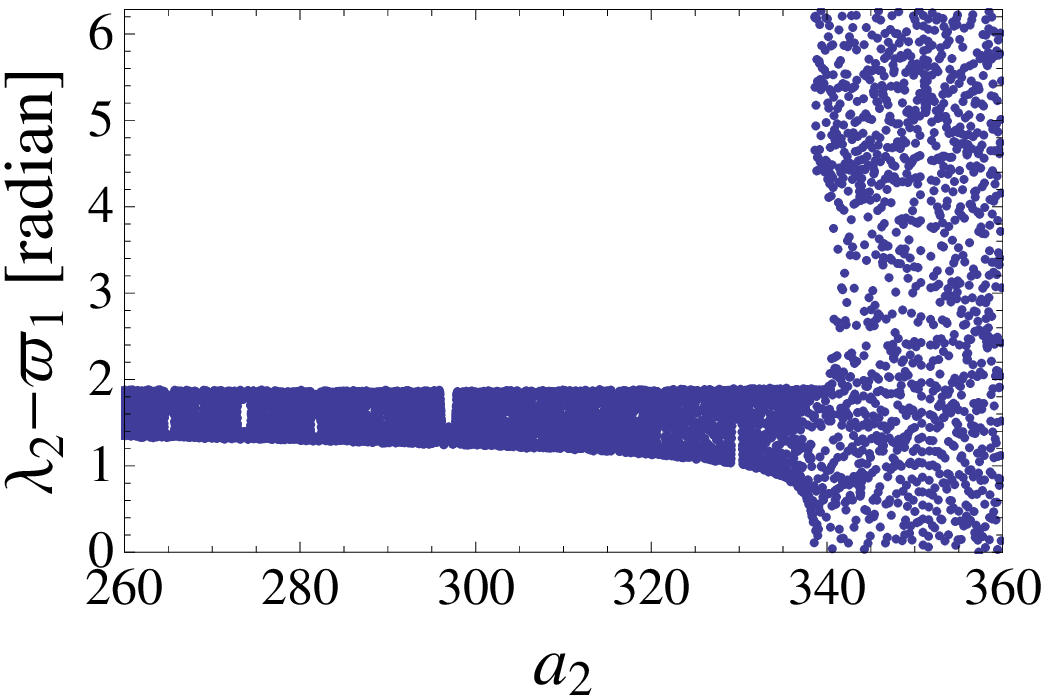}
\includegraphics[width=40mm]{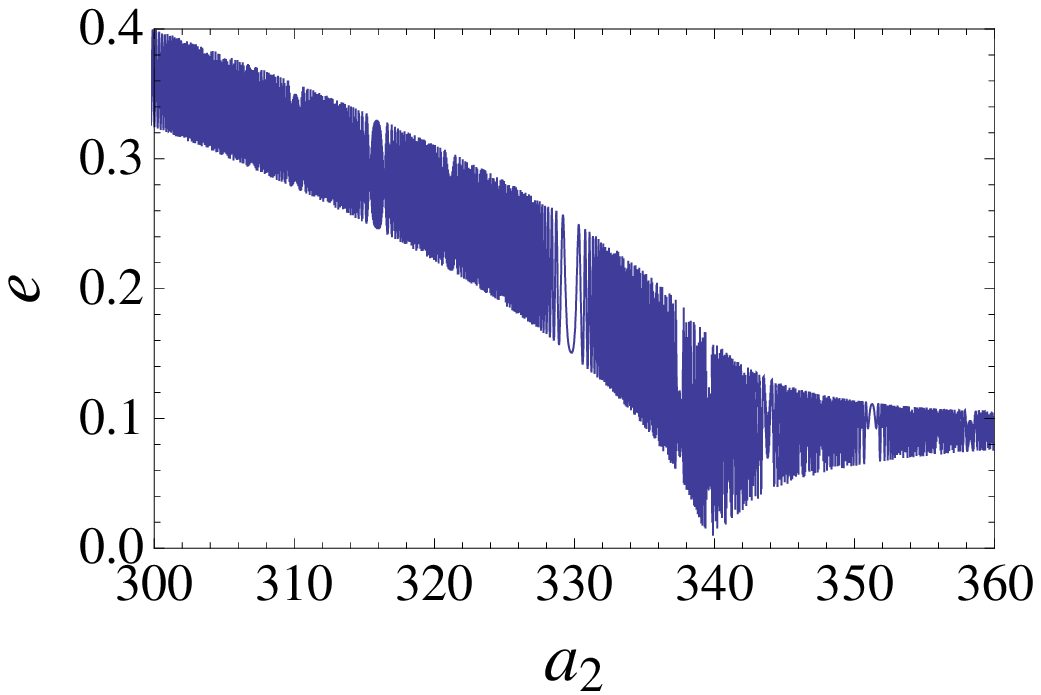}
\includegraphics[width=40mm]{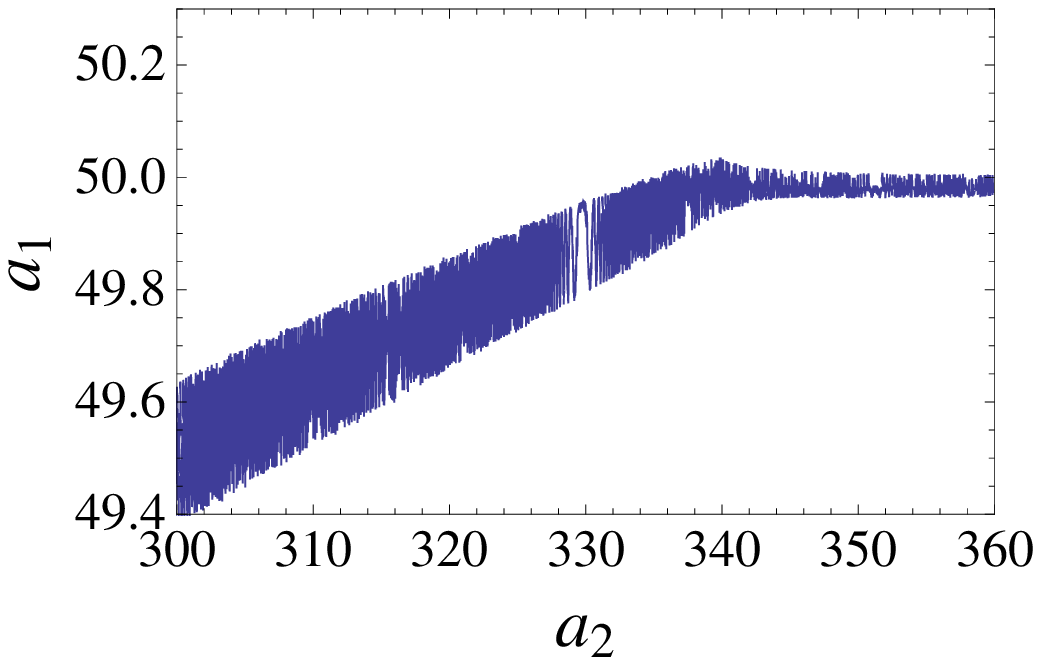}
\includegraphics[width=40mm]{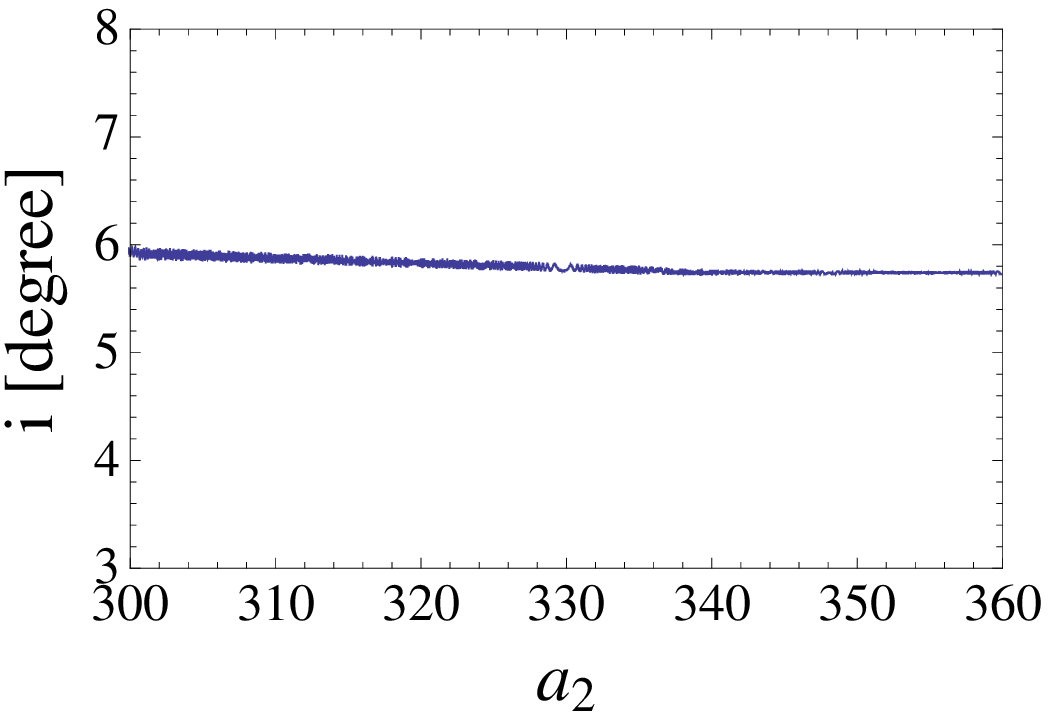}
\caption{Evolution of an EMRI-MBH triple system from an inner eccentricity 
$e=0.10$ (model I). This system has initial parameters similar to Fig.2, but with an 
initial relative inclination $i\sim5.8^\circ$. 
The EMRI is captured into resonance around $a_2\simeq 340$.  The 
inclination angle $i$ is within $5.7^\circ<i<6.0^\circ$. 
}
\label{esx}
\end{figure}

\begin{figure}
\includegraphics[width=40mm]{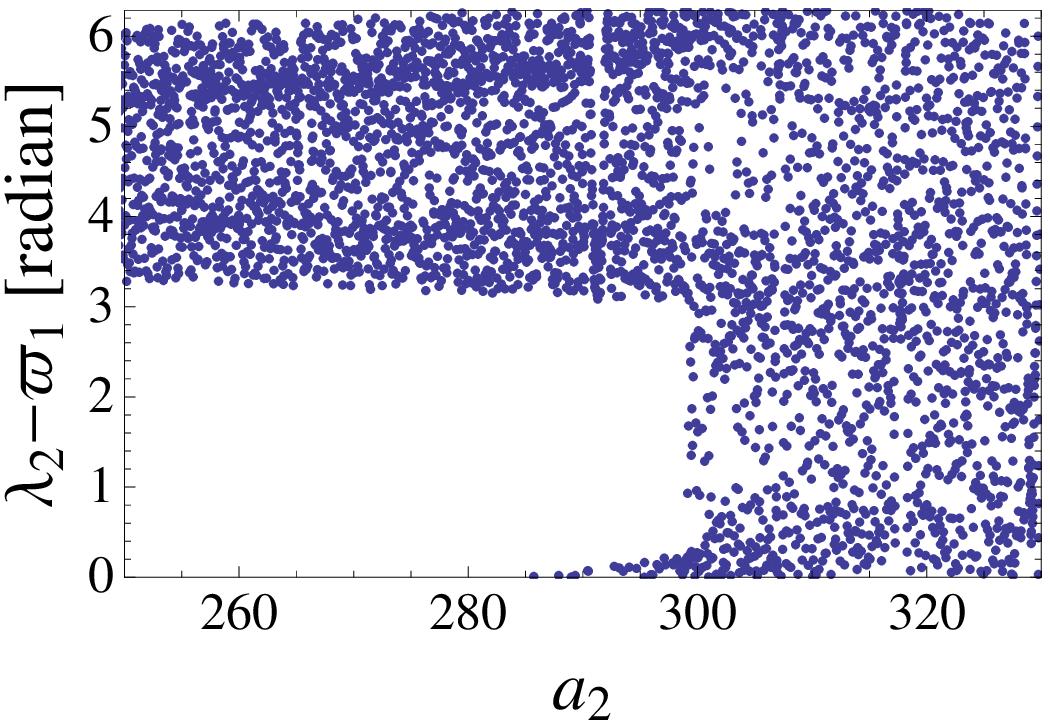}
\includegraphics[width=40mm]{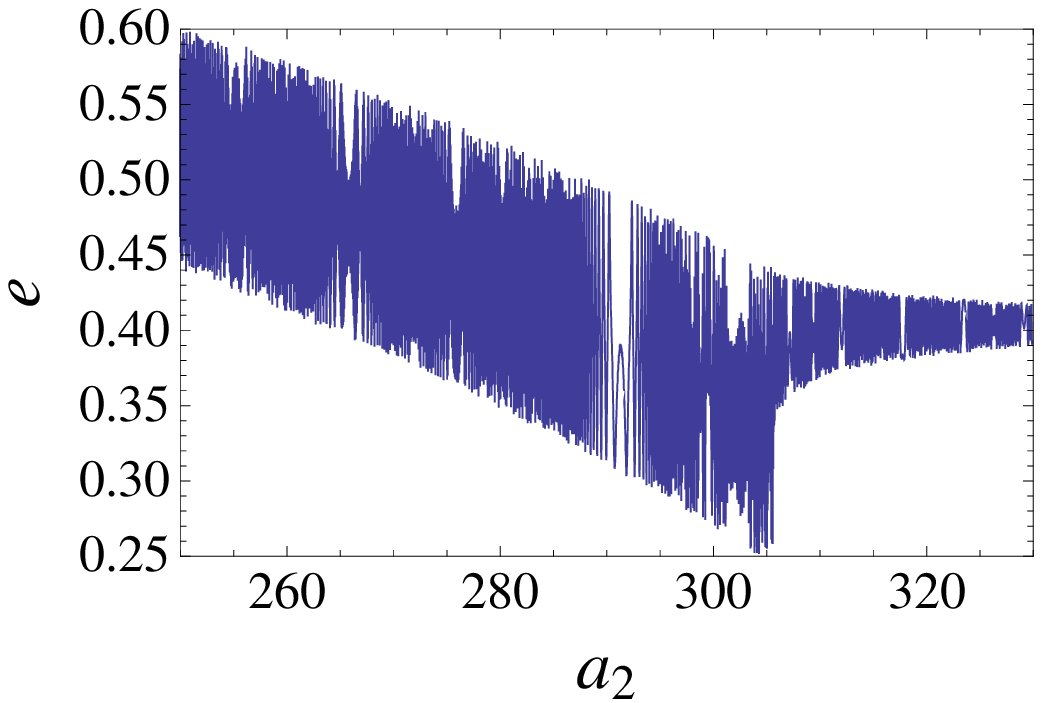}
\includegraphics[width=40mm]{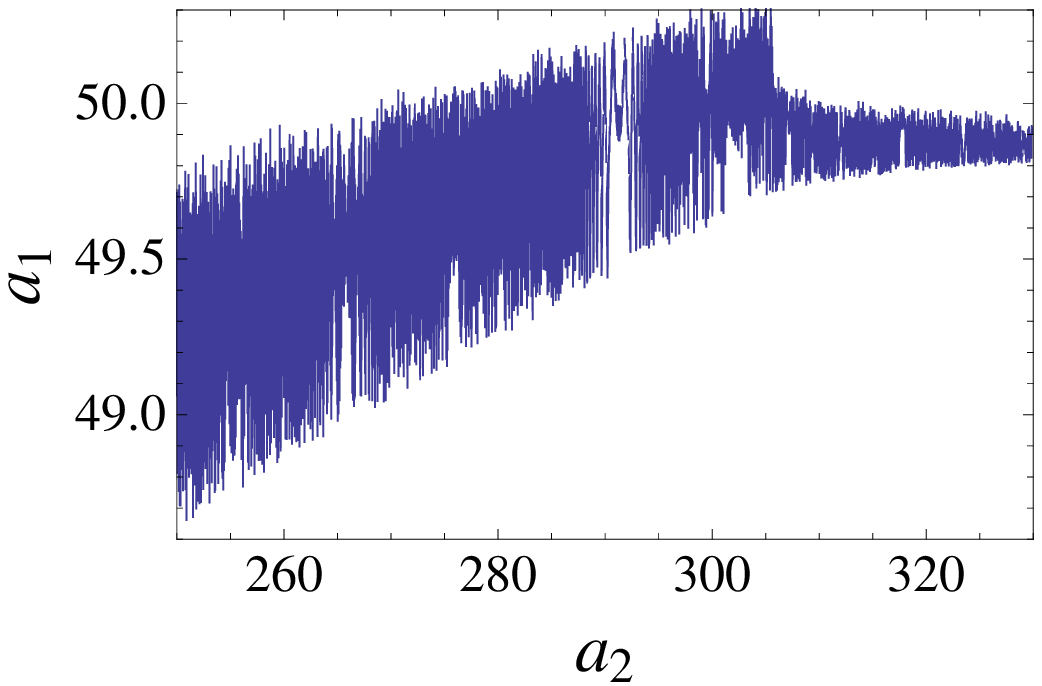}
\includegraphics[width=40mm]{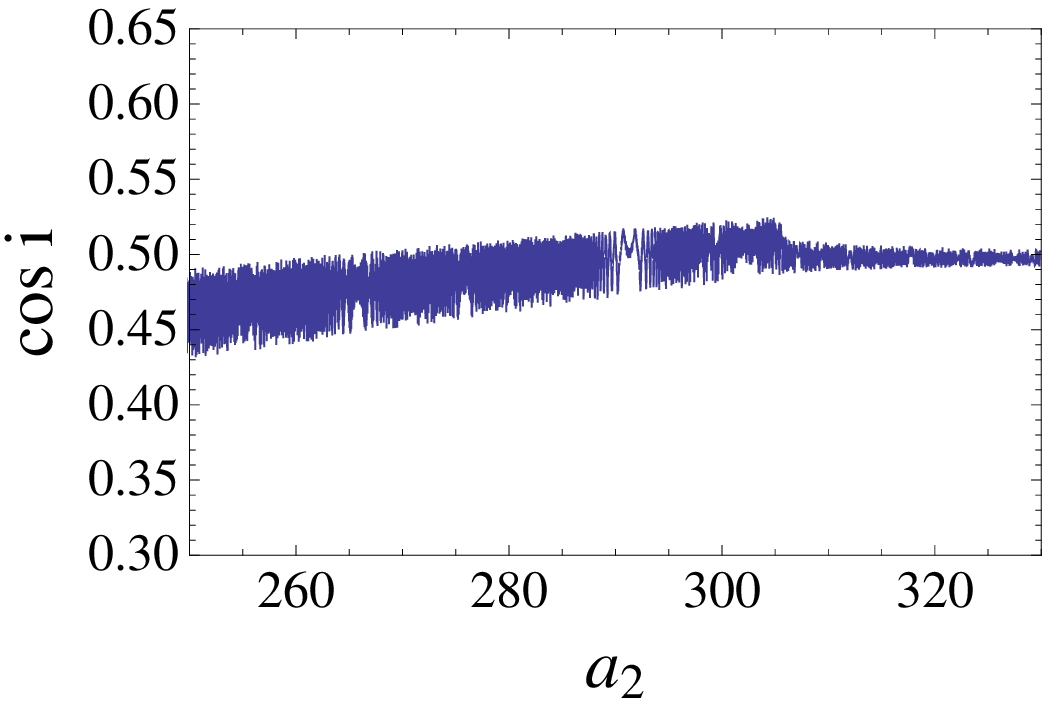}
\caption{Evolution of an EMRI-MBH triple system from an inner eccentricity 
$e\sim0.39$ and relative inclination $i\sim 60^\circ$ (model I).
This system encounters the resonance around $a_2\simeq 305$ where the inner 
eccentricity $e$ shows larger variation $\Delta e\sim 0.2$ compared with  
$\Delta \cos i\sim 0.03$.  
}
\label{esx}
\end{figure}

In the  analytical Hamiltonian 
approach,  we need to clarify how  the resonant interaction depends on the 
inclination angle. { Here, based on the above numerical demonstrations, we 
make  an
approximation that the inclination angle $i$ is constant around the resonant encounter.}
In the previous case for coplanar orbits,  the disturbing function has  the 
 following second-order resonance term 
\beq
\zeta=\frac{15\alpha^2e^2}8\cos2\eta.
\eeq
Here we neglected subleading contributions of $o(\alpha^2e^2)$.
In celestial mechanics,  the effects of the inclination on the disturbing 
function are often handled perturbatively with the expansion parameter 
$s\equiv \sin[i/2]$.  But,  here, we are interested  in highly inclined 
orbits with $s=O(1)$, well beyond the perturbative regime $s\ll 1$.

{ We should notice that the factor ${\bar \epsilon}$ for the resonant 
term in the Hamiltonian (\ref{oh}) is contributed by all the terms proportional to 
$\cos 2\eta$ among the disturbing function.}
Fortunately, we can readily collect  the terms at the lowest order 
$\alpha^2e^2\cos2\eta$ as follows;
\beqa
\zeta&=&\frac{15 \alpha^2 e^2}{8}(1-2s^2+s^4)\cos2\eta\\
  &=&\frac{15 \alpha^2 e^2}{8}\cos^4[i/2]\cos2\eta.
\eeqa

Therefore,  with respect to the original Hamiltonian ${\cal H}^\dagger$ given 
in Eq.(\ref{oh}) and at the order $O(\alpha^2e^2)$ of the resonant interaction,  we just need to multiply the factor $\cos^4(i/2)$ to the parameter 
$\be$ that was defined in Eq.(\ref{c24}) for the coplanar system.
Under the present approximation $i=const$,  this is basically what we need to do for dealing with the inclined orbits.  
Accordingly,  for the rescaled Hamiltonian (\ref{sh2}) in the test particle 
limit (see \S 5.2),  the coefficients $D$ in Eq.(\ref{cd}) and the transition speed 
$d\bd/d\tau$ are given  as
\beq
D=\frac{2a_1}{45m_0m_2\cos^4[i/2]},
\eeq
\beq
\frac{d\bd}{d\tau}=\frac{512m_9^{3/2}}{3^{1/3}375 
m_2 a_1^{13/6}\cos^8[i/2]}.
\eeq
With these expressions, it is straightforward to apply the previous analytical 
methods  in \S 6  and 7  to inclined orbits.

Now, we compare these analytical predictions with numerical simulations 
for triple systems.  Below, we fix the inclination angle at a 
relatively large value $i=60^\circ$ and performed a large number of simulations 
for  two models I and II. 
Even with the strong dependence on the inclination $\propto 1/\cos^8[i/2]=(4/3)^4=3.2$, 
the transition speeds $d\bd/d\tau$ are  less than 0.1 both with models I and II 
(see Table 1), and the adiabatic approximation would be still effective for analyzing
resonant dynamics.

Note also that due to the relativistic apsidal precession, 
the  Kozai process \citep{1962AJ.....67..591K,1962P&SS....9..719L}  does not work here \citep{1997Natur.386..254H,2002ApJ...578..775B,2012PhRvD..85f4037S}.
{ Even for  $m_2\sim m_1$, the characteristic frequency of Kozai process is $O(
\alpha^3 n_1)$, while the 1PN precession frequency of the inner EMRI is  $\sim 3p n_1$. When a system 
encounters our resonance, we have $\alpha^{3/2}=3p\ll 1$ (see  Eq.(\ref{equi})) and the Kozai 
process is suppressed by the 1PN precession effect (see also Naoz et al. 
2012). Note that the semi major axes  of the systems shown here are not constant of motion, and  the orbital averaging  associate with the Kozai mechanism cannot be applied here. These systems lay below the stability criterion presented in Lithwick \& Naoz 2011.

}

 As for the coplanar orbits, we present 
 the capture rates (Fig.12) and the gaps of the eccentricities 
(Fig.13).  The analytical predictions based on the Hamiltonian approach show 
good agreements with the numerical results at $e\lsim 0.4$.  


The favourable results in this section could be regarded as additional supports 
for validity of our  Hamiltonian approach extended for the  relativistic resonance.

\begin{figure}\label{ratei}
  \includegraphics[width=70mm]{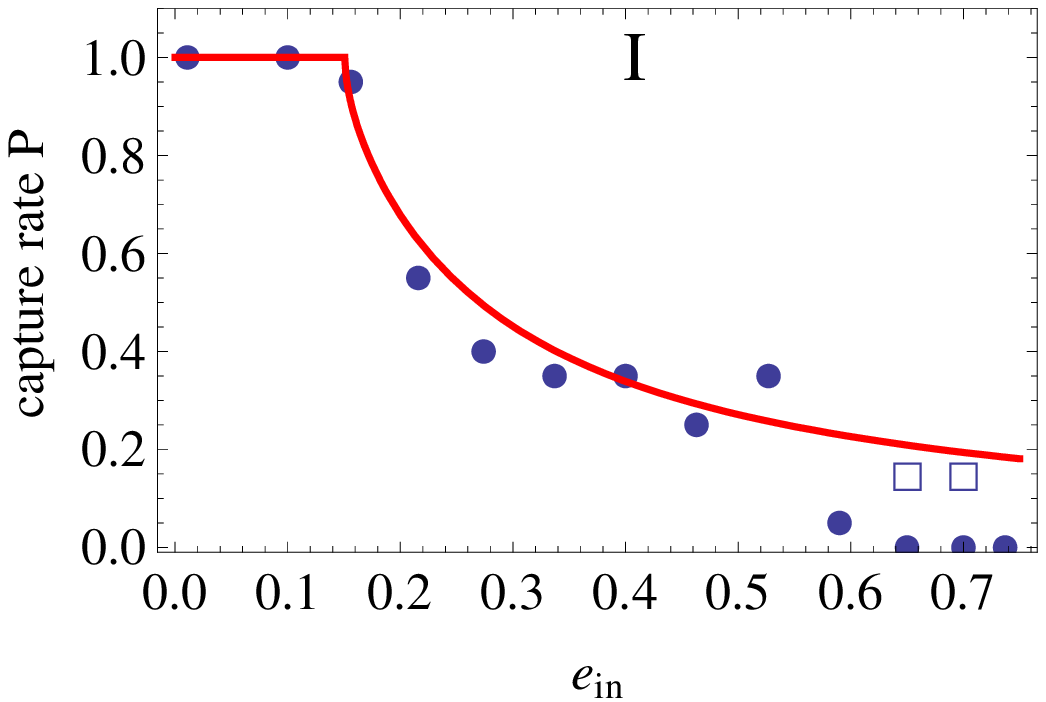} 
  \includegraphics[width=70mm]{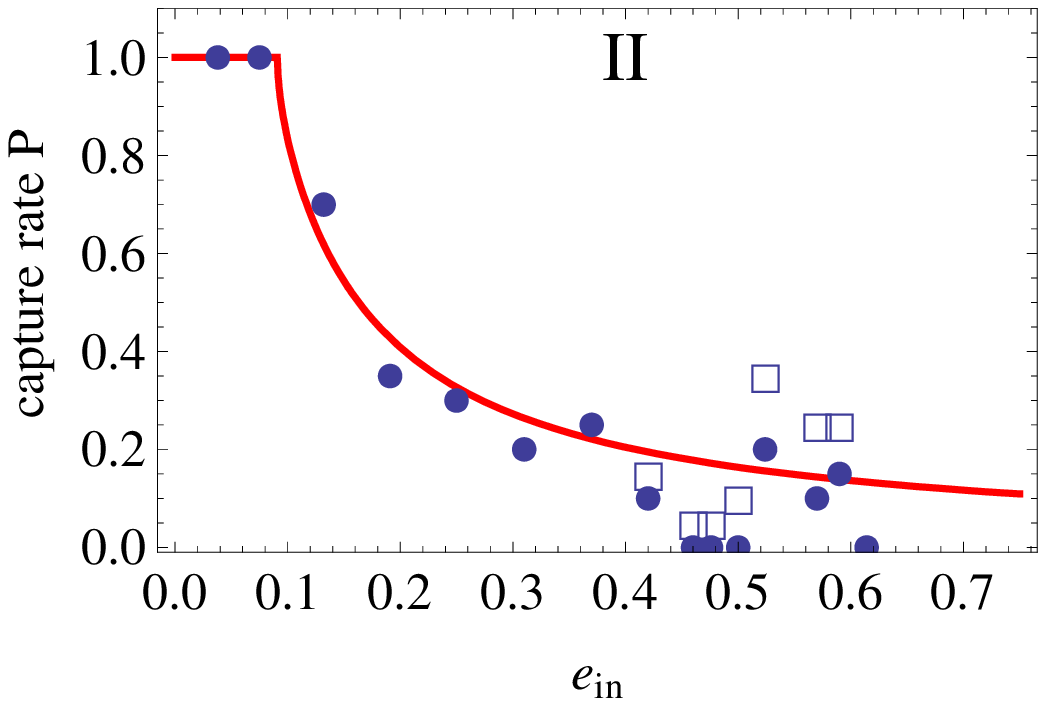}
  \caption{ Resonant capture rate of an inner EMRI  with the inclination angle 
$i=60^\circ$.  The solid curves are given by the Hamiltonian approach with the second 
order mode.  The circles are results from numerical results (each from 20 runs) 
with the
minimum duration $a_2/10$ of the resonant state, and the additional open squares are for $a_2/20$ 
(shown only if different).
 }
\end{figure}

\begin{figure}\label{gapi}
  \includegraphics[width=70mm]{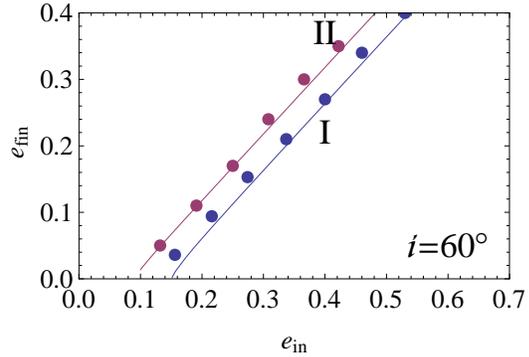} 
  \caption{Correspondences of the eccentricities around the resonant 
encounters. The results are given for models I and II, as in Fig.8, but now for inclined orbits with $i=60^\circ$.
 }
\end{figure}

\section{Summary}

We have studied dynamics of the resonant state $\lambda_2-\varpi_1\sim const$ 
for a triple system composed by  an EMRI (CO+MBH) and an additional outer MBH. 
This resonant state is supported by the relativistic apsidal precession of the 
inner EMRI, and does no depend on its mean anomaly $\lambda_1$.  As a result, 
the two orbits can become hierarchical with $\alpha\equiv a_1/a_2\ll 1$, and 
then the two masses of the MBHs $m_0$ and $m_2$ can become comparable, in 
contrast to the standard MMRs where we have $m_1,m_2\ll m_0$ due to dynamical 
stability of orbits with $\alpha=O(1)$.

As a preliminary analysis, in \S 4, we discussed the dominant order of the 
resonant interaction for our state $\lambda_2-\varpi_1\sim const$.  Due to the 
orbital hierarchy, dependencies on the parameter $\alpha(\ll 1)$ play a critical 
role to assess the relevant terms, and the second-order one $\propto 
\alpha^2e^2$ ($e$: the eccentricity of  the EMRI) could become more important than the 
first-order one $\propto e \alpha^3$. This result is remarkably different from 
the standard MMRs for which the parameter $\alpha=O(1)$ is  less important.

In \S 5, we derive the mapping from the resonant triple systems to the rescaled one-dimensional Hamiltonian for the state  
$\lambda_2-\varpi_1\sim const$. We basically followed the framework of the 
Hamiltonian approach explained  in the literature, but payed special attention to the term 
associated with the relativistic apsidal precession $\dv_{1r}$.   Here the 
dependence $\dv_{1r}\propto (1-e^2)^{-1}$ on the eccentricity $e$ is the key 
element for the structure of the derived Hamiltonian.  The mapping from the 
original EMRI-MBH triple system to the rescaled Hamiltonian becomes regular even 
with the test particle limit $m_1\to 0$ where the difference from the standard 
MMRs would become clear.

Then,  based on the derived mapping, we made analytical predictions on the 
dynamical evolution of the state  $\lambda_2-\varpi_1\sim const$ around the resonant 
encounter and compare them with numerical simulations. 

In \S 6, we studied the 
resonant capture rate as a function of the eccentricity $e$.  For the analytical 
rate, we incorporated the  mapping derived in \S 5 with  the expressions given by Borderies \& Goldreich (1984) for 
the  rescaled Hamiltonian.  We found that our analytical rates show reasonable 
agreements with numerical results for eccentricity $e\lsim 0.4$
 where we can perturbatively deal with the effects of the eccentricity $e$.

In \S 7,  we studied the gap of the inner eccentricity when the capture is failed. 
With the rescaled Hamiltonian, this characteristic phenomena can be understood as a sudden 
change of periodic motion at a separatrix crossing.  We showed that our 
analytical predictions matches  numerical results well.

Finally, in \S 8,  we discussed relatively inclined orbits.  By evaluating  
dependence of the disturbing function on the inclination angle,  we can derive 
the relevant expressions required for the mapping between the inclined triple 
system to the rescaled Hamiltonian. Again,  our analytical prediction reproduces
 numerical results well for $e\lsim 0.4$.

In  this paper,  setting  EMRI-MBH triple systems as our concrete 
astrophysical targets,  we studied the hierarchical 
resonant state $\lambda_2-\varpi_1\sim const$ induced by relativistic apsidal precession.  Similar analyses 
might be useful for purely Newtonian systems such as a planet orbiting around one 
component of  binary stars.  Also in these cases,  the mapping could be well 
behaved in the test particle limit $m_1\to 0$, due to preferred dependencies of 
the apsidal precession rates on the inner eccentricities.

This work was supported by JSPS (20740151, 24540269) and MEXT (24103006).

\appendix

\section{First Order Resonance}
In Eq.(\ref{zeta}) we expand the resonant potential $\zeta$ up to second-order 
in eccentricity $e$.  In \S 5,  we derive the mapping to the simplified Hamiltonian only 
including the second order mode $e^2 \cos2\eta$.  Here we present formulae 
keeping the first-order mode $\propto e\cos\eta$ alone in the test particle 
limit $m_1\to 0$. The basic procedure is the essentially the same as \S 5. After 
some algebra, the rescaled Hamiltonian is given in the forms
\beq
H=\Phi^2+\bd \Phi-\Phi^{1/2}\cos(\phi) \label{sh3}
\eeq
with $\phi\equiv \eta$.  The momentum $\Phi$ is related to eccentricity as
\beq
\Phi=E e^2
\eeq
with the coefficient
\beq
E\equiv \frac{2^{4/3}a_1^{10/9}}{3^{16/9}5^{2/3}m_2^{2/3}m_0^{4/3}}. \label{coe}
\eeq
Meanwhile the transition speed $d\bd/d\tau$ is expressed with the original parameters 
of the triple system as
\beq
 \frac{d\bd}{d\tau}=-\frac{2^{29/3}m_0^{5/6}}{3^{8/9}5^{7/3}a_1^{35/18}m_2^{1/3}}.\label{vel2}
\eeq 
Note that Eqs.(\ref{coe}) and (\ref{vel2}) are regular in the test 
particle limit, as in the case for the  second-order mode.

\section{analytical formulae for the resonant capture rates}
Borderies \&  Goldreich (1984) analytically evaluated the resonant capture rate for a 
simplified Hamiltonian 
\beq
H'=\Phi^2 +\bd~ \Phi+(-1)^k\Phi^{k/2} \cos k\phi \label{b1}
\eeq
with $k=1$ (first-order resonance) and $k=2$ (second-order resonance). Here a 
capture means that the motion of the angle $\phi$ shifts from a 
rotation in the full angular range $[0,2\pi]$ to a libration within a limited 
range.  {  We should note that, as partially demonstrated
 in \S 5, perturbative  expansion for the eccentricity $e$ is crucial to derive 
 the rather simple forms (\ref{b1}).  }

The goal in this appendix is to provide their capture rates as functions of the initial momentum $\Phi_{in}$.   We do not intend to fully explain the arguments in Borderies \& Goldreich (1984), but concisely presents their main results.
We basically deal with the two 
cases $k=2$ and $k=1$ separately in \S B1 and B2 below. Before going into analyses specific to each  mode, we firstly 
describe the features  common to $k=2$ 
and 1.   Here we  omit the label $k$, 
if unnecessary

We consider  a case when the parameter $\bd$  decreases adiabatically from $\bd 
\gg 1$ down to $\bd \ll -1$.
At an initial epoch $\bd_{in}\gg 1$, far from resonance,  the structure 
of a curve $H'=const$ is simple.  We have $\Phi\simeq \Phi_{in}=const$ with 
the  conjugate variable $\phi$ rotating in the full range $[0,2\pi]$. When the 
parameter $\bd$ becomes less than a critical value $\bd_{cr}$ (depending on the 
parameter $k$), 
 the 
Hamiltonian $H'$ now has an inner separatrix, in addition to an outer one.
We denote   the areas inside these two separatrixes by  $2\pi \Phi_I$ (inner) and $2\pi 
\Phi_O$ (outer), as functions of the epoch $\bd$. 
For the critical value $\bd=\bd_{cr}$, we put $\Phi_{cr}\equiv \Phi_{O}(\bd_{cr})$ (with the identity $\Phi_I(\bd_{cr})=0$).

From conservation of an adiabatic invariant \citep{1969mech.book.....L},  the system with an initial value $\Phi_{in}<\Phi_{cr}$ is captured into the resonance at 
$100\%$, namely with the capture rate of $P(\Phi_{in})=1$. In terms of the 
decreasing parameter $\bd$,  this happens at $\bd>\bd_{cr}$.

 But, for  
$\Phi_{in}\ge \Phi_{cr}$, the resonant encounter occurs at $\bd\le \bd_{cr}$  with 
the capture rate 
$P_k(\Phi_{in})<1$ that was estimated using an 
argument based on energy balance (see {\it e.g.} Goldreich \& Peale 1966). If the capture failed, the system starts to 
rotate near the inner separatrix and soon relaxes to a simple rotation state 
with the magnitude $\Phi=\Phi_I$ ($\Phi_I$: evaluated at the resonant encounter).

Below, for $k=1$ and 2, we separately provide 
the formulae that were  given in Borderies \&  Goldreich (1984)  but 
appropriately adjusted  for the specific  forms of our Hamiltonian (\ref{b1}).

\subsection{Second-Order Resonances; $k=2$}

The
critical values are $\bd_{cr}=-1$ and  $\Phi_{cr}=1$. The capture rate for 
$\Phi_{in}>\Phi_{cr}$ is 
\beq
P_2(\Phi_{in})=\frac2{1+\frac{\pi}{2\arcsin[(-\bd)^{-1/2}]}} \label{p2},
\eeq
where the parameter $\bd$ is related to $\Phi_{in}$ as
\beq
\Phi_{in}=\Phi_{O}(\bd) \label{p2d}
\eeq
with the explicit form $\Phi_{O}(\bd) $ as
\beq
\Phi_{O}(\bd)=\frac{(-\bd-1)^{1/2}}\pi  -\frac{\bd}\pi  \lnk \frac{\pi}2+\arcsin[(-\bd)^{-1/2}]  \rnk \label{kai}.
\eeq
The expression for $\Phi_I(\bd)$ is given in a similar complicated form.
But, using a formula of a definite integral, we  can obtain a simple relation 
(Malhotra 1988)
\beq
\Phi_O+\Phi_I=-\bd.
\eeq

\begin{figure}\label{fbg}
  \includegraphics[width=70mm]{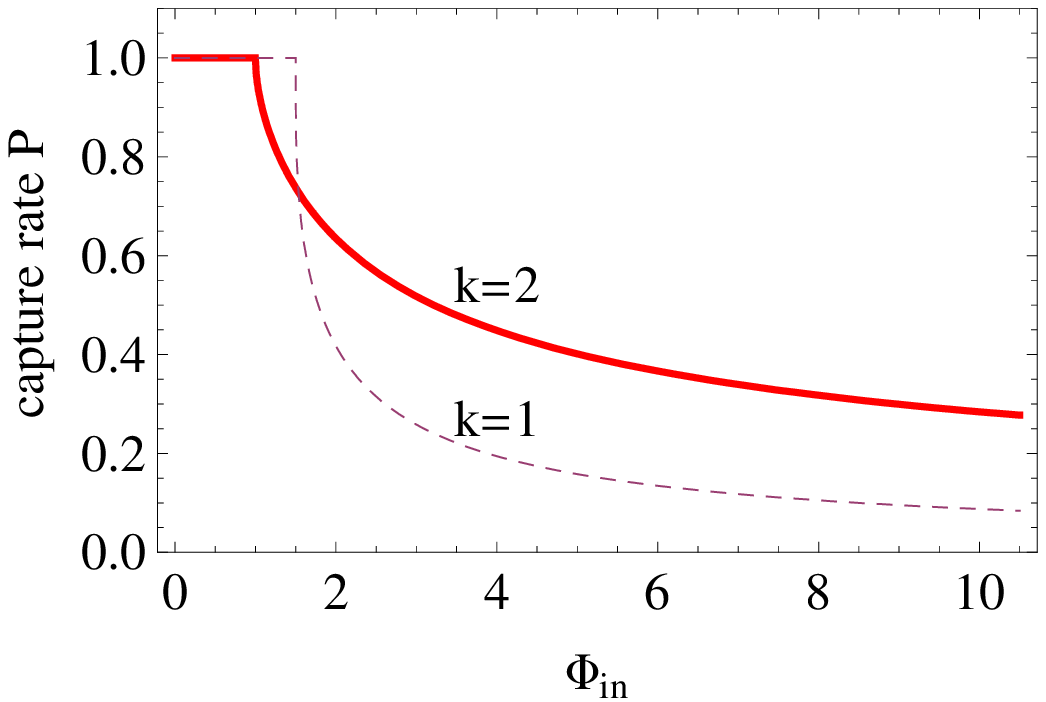} 
  \caption{Resonant capture rates for the scaled Hamiltonian (B1) with $k=1$ 
(first-order resonance) and $k=2$ (second-order resonance).
 }
\end{figure}

We plot the rate for $k=2$ in Fig.\ref{fbg} with the solid curve.

\subsection{First-Order Resonances; $k=1$}
The
critical values are $\bd_{cr}=-3/2$ and  $\Phi_{cr}=3/2$. The rate for 
$\Phi_{in}>\Phi_{cr}$ is given as
\beq
P_1(\Phi_{in})=\frac2{1+\frac{\pi}{2\arcsin[(-2\bd u/3)^{-3/2}]}},
\eeq
where the two parameters $\bd$ and $u$ are the solutions of the following two equations
\beqa
2\pi\Phi_{in}&=&\frac43\bd^2 \lnk \frac{\pi}2+\arcsin\lkk \lmk \frac{-2\bd u}3\rmk^{-3/2}\rkk  
\rnk\nonumber\\
& &-\frac9{2\bd u}\lkk\lmk \frac{-2\bd u}3\rmk^3-1\rkk^{1/2}\nonumber\\
-\frac{27}{4\bd^3u}&=&3-u^2.
\eeqa
In Fig.\ref{fbg} the capture rate for $k=1$ is shown with the dashed curve.

\section{additional canonical transformation}

The rescaled Hamiltonians $H$  in Eqs.(\ref{sh2}) and (\ref{sh3}) are given for 
the variable $\phi$ (closely related to $\eta$) and its conjugate 
momentum $\Phi (\propto e^2)$.  The corresponding canonical equations are 
written as
\beq
\frac{\p \phi}{\p \tau }=\frac{\p H}{\p \Phi},~~\frac{\p \Phi}{\p \tau }=-\frac{\p H}{\p \phi}.
\eeq
In the literature, an additional canonical transformation $(\phi,\Phi)\to (x,y)$ is
frequently introduced as follows
\beq
x=\sqrt{2\Phi}\cos\phi,~~y=\sqrt{2\Phi}\sin\phi.
\eeq
In the new phase space, the angle between the $x$-axis and the point $(x,y)$ is 
identical to the original variable $\phi$, and the distance to the origin is 
proportional to the eccentricity as $\sqrt{x^2+y^2}\propto \sqrt{\Phi}\propto e$. 
With these new variables, the rescaled Hamiltonians $H$ become simple polynomials as
\beq
H=\lmk  \frac{x^2+y^2}2 \rmk^2+{\bd} \frac{x^2+y^2}2+\frac{x^2-y^2}{2}
\eeq
for the second-order one (\ref{sh2}), and 
\beq
H=\lmk  \frac{x^2+y^2}2 \rmk^2+{\bd} \frac{x^2+y^2}2-\frac{x}{\sqrt{2}}
\eeq
for the first-order one (\ref{sh3}).

The canonical variables $(x,y)$ have geometrically intuitive meanings, and, in many cases, 
they  are more convenient to analyze the dynamics of the rescaled Hamiltonians 
themselves, as widely done  in preceding studies (Borderies \&  Goldreich 
1984; Peale 1986; MD).  But we concentrate on the issues more related to the 
mapping between the Hamiltonians and the EMRI-MBH triple systems, and stay  away 
from the new variables $(x,y)$ in the rest of this paper.

\end{document}